\documentclass[fleqn,usenatbib]{mnras}

\usepackage{newtxtext,newtxmath}
\usepackage[T1]{fontenc}

\DeclareRobustCommand{\VAN}[3]{#2}
\let\VANthebibliography\thebibliography
\def\thebibliography{\DeclareRobustCommand{\VAN}[3]{##3}\VANthebibliography}

\usepackage{graphicx}	% Including figure files
\usepackage{amsmath}	% Advanced maths commands

\title[Effect of Stellar Feedback on Gas Stripping]{The Surprising Lack of Effect from Stellar Feedback on the Gas Stripping Rate from Massive Jellyfish Galaxies}

\author[N. Akerman et al.]{
Nina Akerman,$^{1, 2}$\thanks{E-mail: nina.akerman@studenti.unipd.it}
Stephanie Tonnesen,$^{3}$
Bianca Maria Poggianti,$^{1}$
Rory Smith,$^{4}$
Antonino Marasco,$^{1}$
\newauthor
Andrea Kulier,$^{1}$
Ancla M\"{u}ller,$^{5}$
and Benedetta Vulcani,$^{1}$
\\
$^{1}$INAF - Astronomical Observatory of Padova, vicolo dell'Osservatorio 5, IT-35122 Padova, Italy\\
$^{2}$Dipartimento di Fisica e Astronomia `Galileo Galilei', Università di Padova, vicolo dell'Osservatorio 3, IT-35122, Padova, Italy\\
$^{3}$Flatiron Institute, CCA, 162 5th Avenue, New York, NY 10010, USA\\
$^{4}$Departamento de Física, Universidad Técnica Federico Santa María, Vicuña Mackenna 3939, San Joaquín, Santiago de Chile\\
$^{5}$Ruhr University Bochum (RUB), Faculty of Physics and Astronomy, Astronomical Institute, Universitätsstr. 150, 44801 Bochum, Germany
}

\date{Accepted XXX. Received YYY; in original form ZZZ}

\pubyear{2023}

\begin{document}
\label{firstpage}
\pagerange{\pageref{firstpage}--\pageref{lastpage}}
\maketitle

\begin{abstract}
We study the role of star formation and stellar feedback in a galaxy being ram pressure stripped on its infall into a cluster. We use hydrodynamical wind-tunnel simulations of a massive galaxy ($M_\text{star} = 10^{11} M_\odot$) moving into a massive cluster ($M_\text{cluster} = 10^{15} M_\odot$). We have two types of simulations: with and without star formation and stellar feedback, SF and RC respectively. For each type we simulate four realisations of the same galaxy: a face-on wind, edge-on wind, $45^\circ$ angled wind, and a control galaxy not subject to ram pressure. We directly compare the stripping evolution of galaxies with and without star formation. We find that stellar feedback has no direct effect on the stripping process, i.e. there is no enhancement in stripping via a velocity kick to the interstellar medium gas. The main difference between RC and SF galaxies is due to the indirect effect of stellar feedback, which produces a smoother and more homogeneous interstellar medium. Hence, while the average gas surface density is comparable in both simulation types, the scatter is broader in the RC galaxies. As a result, at the galaxy outskirts overdense clumps survive in RC simulation, and the stripping proceeds more slowly. At the same time, in the inner disc, underdense gas in the RC holes is removed faster than the smoothly distributed gas in the SF simulation. For our massive galaxy, we therefore find that the effect of feedback on the stripping rate is almost negligible, independent of wind angle.
\end{abstract}


%%%%%%%%%%%%%%%%%%%%%%%%%%%%%%%%%%%%%%%%%%%%%%%%%%

%%%%%%%%%%%%%%%%% BODY OF PAPER %%%%%%%%%%%%%%%%%%

\section{Introduction} \label{sec:intro}

A cluster galaxy may be subject to processes specific to high-density environments. Among them is ram pressure stripping \citep[RPS;][]{GG72} in which the intracluster medium (ICM) interacts with the interstellar medium (ISM) of the galaxy, removing it as the galaxy falls into the cluster potential well. As a result, over time the galaxy quenches its star formation, while the stripped gas forms tails trailing behind the galaxy disc \citep{Gavazzi01, Sun10, Sun22, Fumagalli14, Kenney14, Fossati16, Jachym17, Jachym19, George018, GASPXXII}. Observational evidence \citep[corroborated by simple theoretical predictions, e.g.][]{GG72} suggests that as the gas is stripped outside-in \citep{Cortese12, Merluzzi16, GASPIV, Fossati18, Cramer2019, GASPXXIV} and the disc is left truncated with a disturbed morphology \citep{Koopmann&Kenney04, Crowl&Kenney2008, Boselli14, GASPIII, Kenney15}. The effects outlined above have long been observed in HI gas \citep{Cayatte1990, Kenney04, Chung07}, which can be ionised by in-situ star formation or other processes, giving rise to H$\alpha$ emitting-tails \citep{Fumagalli14, Merluzzi16, GASPI, GASPII, GASPIV}. Recent observations showed that such tails also contain large amounts of molecular gas, partly formed in-situ and partly stripped in the vicinity of the disc \citep{Lee17, Jachym17, Jachym19, Lee&Chung18, GASPX, GASPXXII, Moretti20}.

Galaxy evolution under RPS has also been studied in cosmological and idealised simulations. Some of the first simulations used a constant ICM wind \citep{SchulzStruck01, Roediger&Bruggen06, Tonnesen09}, though \cite{Tonnesen19} showed that modelling ram pressure (RP) whose magnitude increases with time (as experienced by a galaxy falling into a cluster) is crucial for obtaining realistic stripping profiles of galaxies that can be compared to the observed ones. Different angles at which the wind hits the galaxy have also been modelled \citep{Roediger&Bruggen06, Roediger07, Jachym07, Bekki14, Steinhauser16, Akerman23}, with a conclusion that there is a weak dependence of the stripping process on the inclination angle unless it is close to edge-on stripping. There are also some simulations that include a magnetised ISM \citep{Dursi&Pfrommer08, Ruszkowski14, Tonnesen&Stone14, Ramos-Martinez18} and cosmic rays \citep{Farber22}, which generally find small changes to the global stripping rate but more significant differences in the gas phase distribution.

Whether RP immediately quenches a galaxy or is able to cause some star formation enhancement via gas compression has long been debated. The latest observations show that some enhancement can indeed take place \citep{Vulcani18, Roberts&Parker20, Roberts21, Roberts22a, Roberts22b}. Simulations agree as well, noting that the star formation rate (SFR) would be increased only temporarily and that whether or not the SFR enhancement would happen depends on many factors, such as galaxy mass, wind inclination angle and RP strength \citep{Tonnesen12, Bekki14, Roediger14, Steinhauser16, Ruggiero&LimaNeto17, Lee20}. \cite{Goller23} find similar results in a large cosmological simulation (TNG50), where SFR is enhanced in individual stripped galaxies at some point of their evolution, although as a population RPS galaxies have low SFRs.

Although simulations have extensively studied the effect of RPS on the SFR, it is still unclear how, in turn, star formation and stellar feedback (hereafter just `feedback' since this is the only type of feedback that we model) influence gas stripping. \cite{Bahe&McCarthy15} show that feedback assists RP in removing gas from a galaxy, with the effect being stronger for low-mass galaxies with $M_\text{star} = (2-5) \times 10^{9} M_\odot$. \cite{Kulier23} also find that in the EAGLE simulation \citep{Schaye15} stripping proceeds quicker due to the stellar feedback, and they note that the feedback might be overestimated in EAGLE \citep[see also][]{Bahe16} due to the subgrid implementation and cosmological-scale resolution.

In this work, we study how star formation and, importantly, stellar feedback, affect the process of RPS by comparing gas in galaxies from two sets of simulations, a first set that includes radiative cooling, and a second set that adds star formation and feedback prescriptions to the radiative cooling simulations. We can naively imagine two scenarios. In the first one, feedback might impede gas stripping by adding more thermal and nonthermal pressure support to and above the ISM that will block the RP from making direct contact with the disc (this will be most effective when the wind has a significant component in the face-on direction). On the other hand, feedback might enhance gas removal by moving ISM gas higher above and below the disc where it is less tightly gravitationally bound. In this work we find that the impact of feedback is more nuanced than either of the above simple sketches.

The paper is organised as follows. First, we outline our simulations in Section \ref{sec:methods}. In Section \ref{sec:strip_rate} we calculate and compare the stripping rates of star-forming versus radiative-cooling-only galaxies, and in Section \ref{sec:mechanism} we explore what drives the differences in RPS in these sets of simulations. We compare our results to other works in Section \ref{sec:discussion} and draw conclusions in Section \ref{sec:conclusions}.

\section{Methodology} \label{sec:methods}

We describe our simulations in detail in our previous work, \cite{Akerman23}, and here will briefly highlight their main features. Using the adaptive mesh refinement code Enzo \citep{Enzo}, we make a simulation box with 160 kpc on a side, with 5 levels of refinement allowing for a maximum resolution of 39 pc. The refinement criteria are Jeans length and baryon mass of $\approx7500 M_\odot$. In Appendix \ref{appendix:refinement} we perform a resolution test and show how our refinement criteria resolve the galaxy disc.

The simulations include radiative cooling using the {\sc grackle} library \citep{Smith17} with metal cooling and the UV background by Haardt \& Madau \citep{Haard&Madaut12}. The disc starts with metallicity $Z = 1.0 \; Z_\odot$ and the ICM has $Z = 0.3 \; Z_\odot$.

We use star formation and stellar feedback recipes by \cite{Goldbaum15, Goldbaum16}. If the mass of a cell exceeds the Jeans mass and the minimum threshold number density of $n_{\rm{thresh}}=10 \, \rm{cm}^{-3}$, a stellar particle will form with a minimum mass of $10^3 M_\odot$, assuming a star formation efficiency of 1 per cent. We also simulated galaxies from the same initial conditions (described below) with $n_{\rm{thresh}} = 3 \, \rm{cm}^{-3}$ and $30 \, \rm{cm}^{-3}$ and found that galaxies with all three $n_{\rm{thresh}}$ evolve similarly in their star formation history. We study the effect that the choice of the threshold number density might have on our results in Appendix \ref{appendix:star-formation}, where we also outline the role of star formation itself in shaping the ISM distribution. Thus, the parameter choice for the $n_{\rm{thresh}}$ does not notably affect the results. The feedback includes momentum and energy input from supernovae (SNe) (that also increase cell metallicity), ionising radiation from young stars (heating up to $10^4$ K), and winds from evolved massive stars. SNe have combined energy budget of $10^{51}$ erg; first the terminal momentum input is added from the number of SNe expected at a given time step, then any additional energy is added as thermal energy. The total momentum of $3\times10^5 \text{ M}_\odot \text{ km } \text{s}^{-1}$ is distributed equally among the 26 nearest neighbour cells. The maximum change in velocity of any given cell is limited to 1000 km s$^{-1}$.

We follow the setup by \cite{Roediger&Bruggen06} and \cite{Tonnesen09}, modelling static potentials for the stellar disc and the spherical dark matter halo, while calculating the self-gravity of the gas component at each time step. For the Plummer–Kuzmin stellar disc \citep{Miyamoto&Nagai1975} we use a mass of $M_\text{star} = 10^{11} M_\odot$, a scale length of $r_\text{star} = 5.94$ kpc and a scale height of $z_\text{star} = 0.58$ kpc. We model a Burkert profile for the dark matter halo \citep{Burkert1995, Mori&Burkert00} with a core radius of $r_\text{DM} = 17.36$ kpc\footnote{While an Navarro–Frenk–White dark matter potential has been found by many to be more appropriate for massive galaxies \citep[e.g.][]{Cintio14}, a Burkert profile has been found to be a better match to some observational samples of massive galaxies \citep[e.g.][]{Rodrigues17, Li20}. Importantly for this work, the Burkert potential we use leads to a good match between the rotation curve of our simulated galaxy and JO201.  To orient the reader, the R$_{200}$ and M$_{200}$ of our halo are 340 kpc and 1.15$\times$10$^{12}$ M$_{\odot}$, respectively.}. Initial parameters of the gaseous disc are the following: mass $M_\text{gas} = 10^{10} M_\odot$, scale length $r_\text{gas} = 10.1$ kpc and scale height $z_\text{gas} = 0.97$ kpc. The values were chosen to match a well-studied RPS galaxy, JO201 \citep{GASPII, GASPXV}. As with the previous work, our intention here is to merely select realistic initial conditions based on an observed stripped galaxy, not to model it exactly. A more detailed comparison between our simulated galaxies and JO201 can be found in \cite{Akerman23}, Section 7.1.

We fix the galaxy in the centre of the simulation box and add an ICM wind via inflow boundary conditions (and outflow on the opposite side) to simulate RPS. We include a time-varying (in density and velocity) ICM wind \citep{Tonnesen19}, and to find its parameters, we model a galaxy falling into a massive cluster ($M_\text{cluster} = 10^{15} M_\odot$) following the procedure described in \cite{GASPXV} from a clustercentric radius of 1.9 Mpc and with an initial velocity of $1785 \; \mathrm {km \, s}^{-1}$. We assume hydrostatic equilibrium of an isothermal ICM with constant temperature of $T=7.55\times10^7$ K and a beta profile. From Rankine–Hugoniot jump conditions for Mach number of 3 we find the pre-wind ICM conditions using the ICM wind parameters at the initial radius. 

Since our galaxy is fixed in space, we define a wind angle as the angle between the wind direction and the galaxy rotation axis, and model three wind angles: $0^\circ$ (a face-on wind that flows along the $z$-direction in our simulated box, W0), $45^\circ$ (W45, in which the wind has equal components along the $y$- and $z$-directions), and $90^\circ$ (edge-on wind that flows along the $y$-direction in our simulated box, W90). The wind angle is constant throughout a simulation. As a control, we simulate a galaxy that does not undergo RPS (no wind, NW).

We simulate galaxies that include only radiative cooling (denoted as RC) and those that also include star formation and feedback (denoted as SF), thus modelling in total eight galaxies: RCNW, RCW0, RCW45, RCW90 and SFNW, SFW0, SFW45, SFW90.

We evolve our RC galaxies in isolation for 130 Myr, during which the gas cools down and collapses into a thin disc. The wind starts inflowing through the lower boundary of the simulation box, and we restart our simulation as four separate runs: RCNW, RCW0, RCW45, RCW90. The wind reaches the galaxies after 70 Myr, taking 200 Myr in total for the galaxies to evolve before the onset of RP. This amount of time allowed the majority of the gas in the disc (particularly within the stripping radius of 15 kpc, discussed below) to fragment into clouds. We repeat the procedure for the SF galaxies, allowing them to evolve for 230 Myr in isolation in order to stabilise the SFR, such that the variation of the SFR on a 5 Myr time-scale decreases to 5 per cent. In total, the four galaxies (SFNW, SFW0, SFW45, SFW90) evolve for 300 Myr before the beginning of RPS.

The RC simulations run for a total of 900 Myr and the SF simulations for a full Gyr. We used 3.0M node-hours to simulate the four RC galaxies and 3.7M node-hours for the four SF galaxies.

\section{Stripping rate} \label{sec:strip_rate}

\begin{figure}
\centering
\includegraphics[width = 80mm]{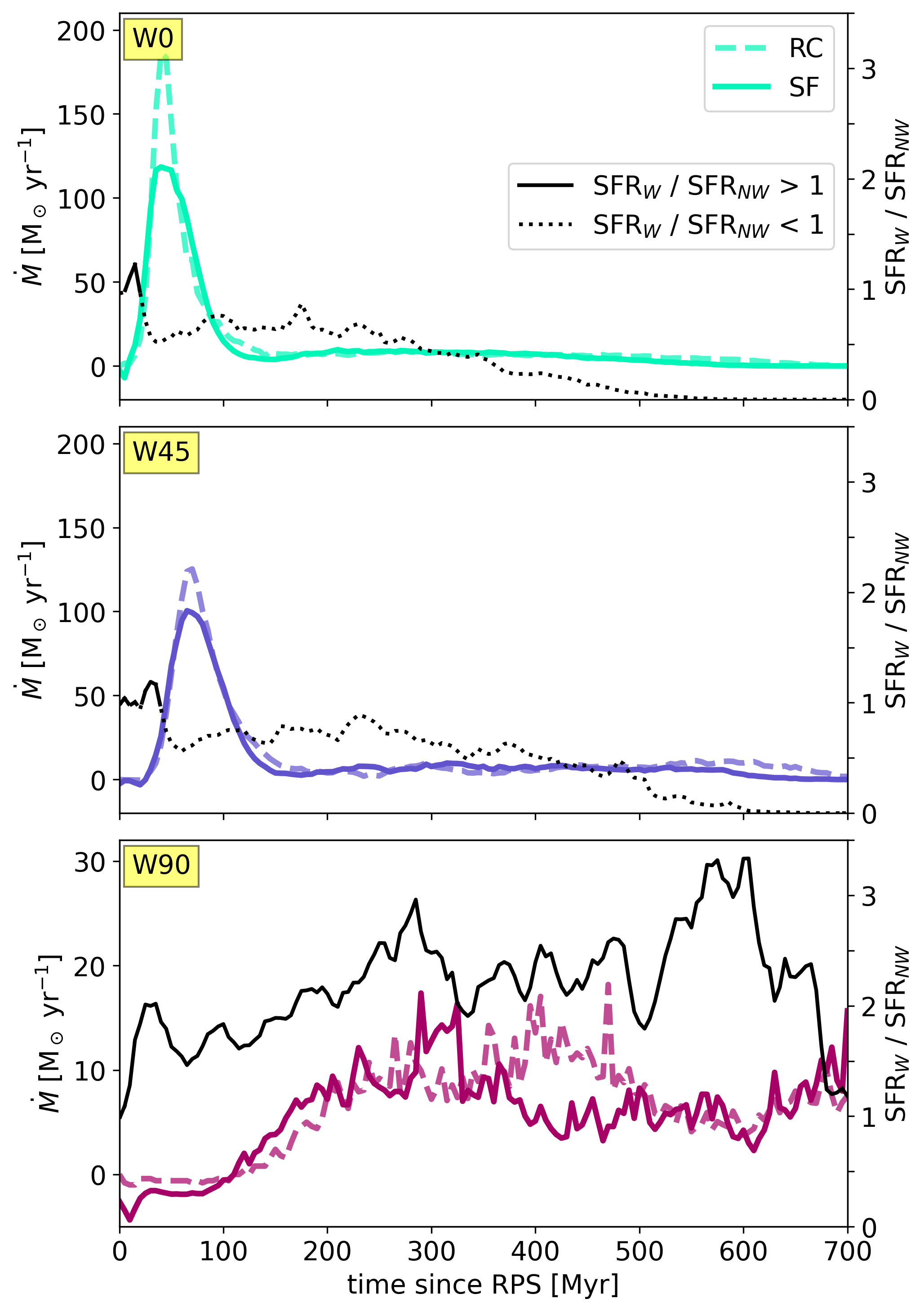}
\caption{Stripping rate of gas with metallicity $Z > 0.35 Z_\odot$ in a galaxy disc (defined with a radius 30 kpc and a height of $\pm2$ kpc from the disc plane) as a function of time. From top to bottom, rows show data for different wind angles, where W0 is a face-on stripped galaxy and W90 is stripped edge-on. Each panel shows $\dot{M}$ for SF (solid) and RC (dashed) simulations. Overplotted in black is the SFR for each galaxy normalised by the SFR in SFNW at the same time step (right $y$-axis), with a solid (dotted) line plotting the ratio above (below) one. Notice that for W0 and W45 (top and middle panels) the left $y$-axis range is the same, while for W90 (bottom panel) it is smaller. The right $y$-axis range is the same in all the panels.}
\label{fig:strip_rate_sf}
\end{figure}

We start by analysing differences in the gas distribution in the galactic disc, a region defined as having a radius of $R=30$ kpc and $h=\pm2$ kpc height from the plane of the disc (we note that we also used a disc with $h=\pm5$ kpc height and found similar results). To quantify the effect that the feedback has on the process of stripping we calculate the stripping rate:

\begin{equation}\label{eqn:striprate}
    \dot{M} = \frac{\Delta M}{\Delta t} - \text{SFR},
\end{equation}

where $\Delta t = t_i - t_{i-1}= 5$ Myr time step, $\Delta M = M_i - M_{i-1}$ is mass difference of gas with metallicity $Z > 0.35 Z_\odot$. The metallicity cut allows us to exclude the pure ICM, and the choice of this metallicity threshold does not have a significant impact on our results, as we tested up to a much more strict metallicity criterion of $0.7Z_\odot$. While equation \ref{eqn:striprate} could be positive in the case of accretion from galaxy fountain flows or the surrounding ICM, once the ICM wind begins removing gas from the galaxy (shortly after RPS begins) $\dot{M}$ is always negative. Because we are examining our simulations while stripping is occurring, we call $\dot{M}$ the stripping rate. SFR is found at the same time step:

\begin{equation}
    \text{SFR} = \frac{M_\text{star}^i}{\Delta t},
\end{equation}

where $M_\text{star}^i$ is the total mass of stars born within the last time step $\Delta t$, which is always 5 Myr. For RC galaxies, SFR is defined to be zero. By subtracting SFR from the stripping rate we account for the gas lost to star formation, thus the differences between stripping rates of RC and SF galaxies should be due to the effect the feedback plays in RPS.

\begin{figure}
\centering
\includegraphics[width = 80mm]{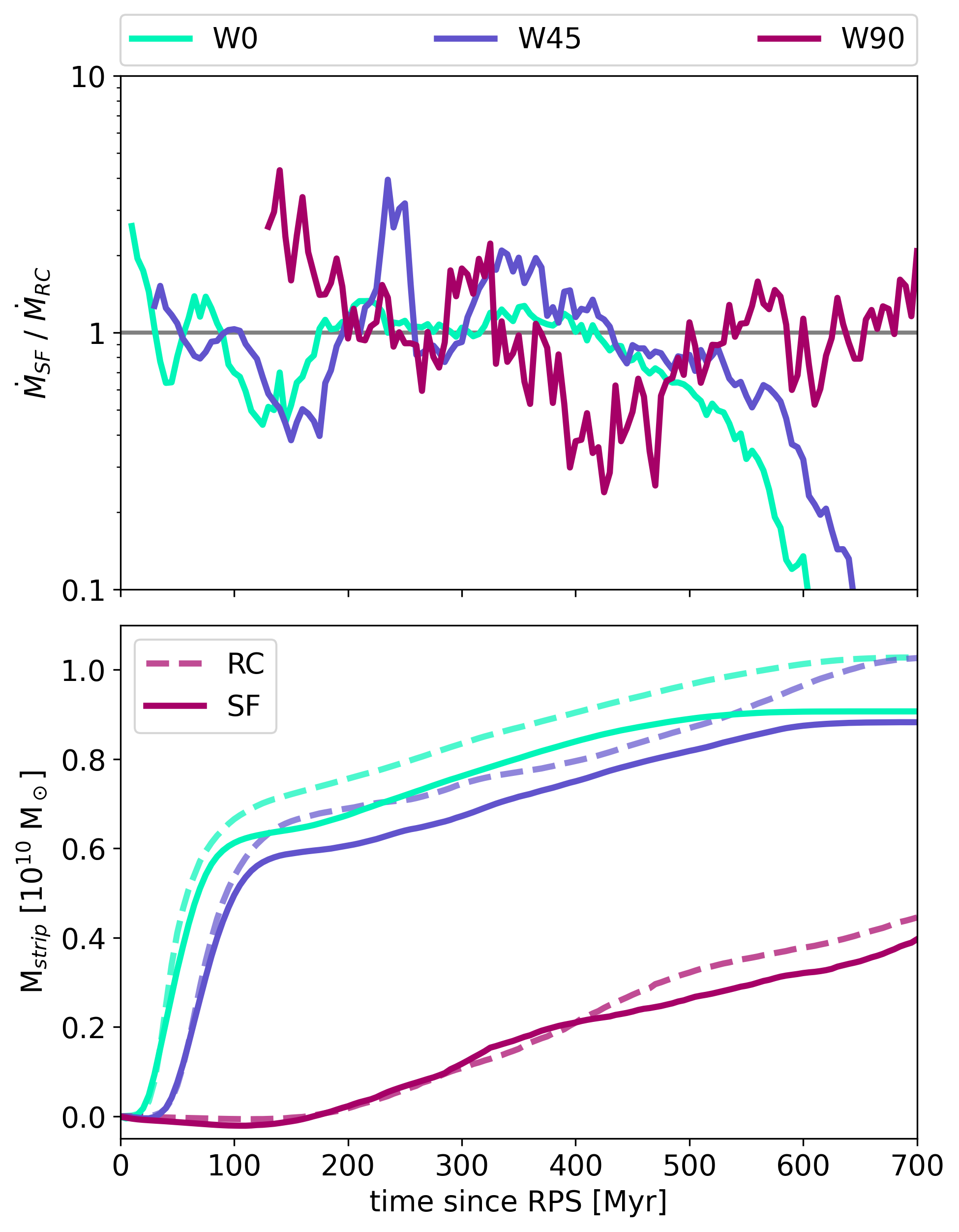}
\caption{{\it Top}: SF to RC stripping rate (from Fig. \ref{fig:strip_rate_sf}) ratio as a function of time, colour-coded by wind angle. The ratio is calculated for $\dot{M}<0$ only, to avoid confusion when gas is accreted at the beginning of the simulation. The grey line of equal ratio serves to guide the eye. The W90 ratio starts at $\sim150$ Myr because before that time the gas is not stripped but accreted onto both of the galaxies. {\it Bottom}: Cumulative mass of stripped gas (accounting for star formation) with metallicity $Z > 0.35 Z_\odot$ as a function of time, for SF (solid) and RC (dashed) simulations.}
\label{fig:strip_mass}
\end{figure}

Fig. \ref{fig:strip_rate_sf} shows the stripping rate, with each row for a different wind angle (W0 is a face-on stripped galaxy and W90 is stripped edge-on), solid lines plotting SF galaxies and dashed lines plotting the RC ones. Note that here and in other plots the $x$-axis (and later any time measurements) is `time since RPS', where $t=0$ denotes the start of RPS and excludes the 200 and 300 Myr during which RC and SF galaxies, respectively, were evolving in isolation. We also confirm that by the start of RPS the SF galaxy has only 5 per cent less gas mass than the RC one, due to star formation, meaning that any variation in the stripping process is not a result of a simple mass difference. We overplot in black the SFR for each galaxy divided by the SFR in SFNW at the same simulation time, with a solid (dotted) line indicating that the ratio is above (below) one. 

Once we have accounted for the gas loss due to star formation, the stripping rates of RC and SF galaxies are very similar for any given wind angle. This might be expected in SFW0 and SFW45 galaxies, since they slowly quench their star formation after the onset of RPS due to the removal of gas in the outskirts, following a brief slight enhancement of SFR during the initial disc compression by the wind. Hence, the global role of feedback in these galaxies will be decreasing with time, and they would become more and more similar to their RC counterparts. Moreover, SFW0 and SFW45 (as well as RCW0 and RCW45) are similar to each other. In \cite{Akerman23} we show that under RP, SFW0 and SFW45 evolve rather closely on the galaxy scale in terms of their gas mass and quenching histories, so their comparable stripping rates (and the striking difference from SFW90) are no surprise. Note that the small delay in the peak stripping rates between W0 and W45 galaxies is simply due to the fact that it takes the angled wind more time to reach the galaxy since it has a larger distance to travel to the centre of the simulation box. At the peak stripping rate of W0 and W45 (at $\sim$30--50 Myr) the RC galaxies are stripped more rapidly than the SF galaxies, indicating that the small SFRs in our W0 and W45 simulations do not enhance stripping. We also note that the W0 maxima are larger than those in W45, suggesting that more angled winds may strip galaxies slower \citep[in agreement with previous works such as][]{Roediger&Bruggen06}. It is more surprising to see that even in SFW90, where the SFR is increased by a factor of 2--3 due to RP, the feedback has almost no apparent effect on the stripping rate. On the contrary, from 350--500 Myr after stripping has begun RCW90 is stripped faster. We will go into the details of where RPS removes mass from the galaxies in the next section. 

To emphasise the variations in the stripping rates in the top panel of Fig. \ref{fig:strip_mass} we plot their ratio, $\dot{M}_\text{SF}$ over $\dot{M}_\text{RC}$. The ratio is calculated for $\dot{M}<0$ only (i.e. outflow), to avoid confusion when gas is accreted at the beginning of the simulation, as the ICM wind starts mixing with the ISM prior to the gas removal. For W0 and W45 the ratio is quite unstable at first, but between 100 to 200 Myr the RC galaxies are stripped systematically faster. The apparent delay in the ratios of W0 and W45 is, again, due to the delay in stripping of the W45 galaxies. Finally, at 200 Myr the W0 galaxies reach the point of `stripping equilibrium' ($\dot{M}_\text{SF}/\dot{M}_\text{RC}\approx1$) that lasts until 400 Myr when $\geq 80$ per cent of these galaxies' original gas has been stripped or used for star formation. The W45 galaxies during most of this period have a higher ratio than that of the W0 at identical times, and also higher than one. However, the stripping rate itself is quite low in all the four galaxies, with $\dot{M}\approx7 M_\odot \text{yr}^{-1}$ on average, which means that even small differences in the stripping rate will be reflected in the ratio.

The differences in the stripping rates of SFW90 and RCW90 show a different pattern, with a more clear trend in the ratio. We note that at the earliest times we do not plot the ratio in the top panel because at the leading edge of the galaxy, where the disc is hit by the wind, the ICM begins mixing with the ISM at a rate higher than that of the gas removal on the opposite side of the galaxy. Because of this, for the first 120 Myr the gas is not stripped but accreted onto both of the galaxies. In RCW90 the total mass of the gas accreted over this period equates to $6.5\times10^7 M_\odot$ or 0.67 per cent of the galaxy mass at the start of RP, while in SFW90 it is $2\times10^8 M_\odot$ or 2 per cent of the galaxy mass. This accretion does not affect the course of stripping. When the gas starts being removed at a steady rate in both SFW90 and RCW90, the stripping rate is higher in the SF galaxy. The ratio then starts gradually but steadily decreasing until 400 Myr, after which it begins rising again.

The bottom panel in Fig. \ref{fig:strip_mass} shows the cumulative amount of stripped gas (this is the integral of Fig. \ref{fig:strip_rate_sf}), i.e. the gas that was removed from a galaxy by RPS, not the total gas mass lost by a galaxy (which would also include gas used in star formation). This distinction is important to keep in mind, since in the figure SFW0 and SFW45 have seemingly lost 15 per cent less gas than the respective RC galaxies by the end of the simulation. This is due to the fact that less gas was available to be stripped from these galaxies, because it was used for star formation, and due to the difference in the stripping rates during 30--50 Myr (to be discussed in Section \ref{sec:mechanism}). In combination, the two panels in the figure illustrate that even though the SF-to-RC stripping rate ratio in the W90 galaxies varies more in the first 500 Myr of RPS, while the ratio in the W0 and W45 stays, in comparison, more constant, the actual mass of stripped gas in the W90 galaxies is much lower. As also shown in Fig. \ref{fig:strip_rate_sf}, before 200 Myr, the stripping rate in the W0 and W45 galaxies is much higher than in the W90 galaxy, and even small disagreements between the RC and SF stripping rates, when accumulated, lead to noticeable differences in the amount of stripped gas. Conversely, RCW90 and SFW90 lose much less gas, especially in the initial 100 Myr.

In none of the three galaxies can the behaviour of the stripping rate ratio be easily tied to the presence of feedback. One could imagine that it would either assist or impede the RPS. By expelling the gas high above the plane of the disc (up to 15 kpc, see below), the feedback makes the gas less tightly gravitationally bound and thus more easily stripped. On the other hand, the feedback adds thermal and nonthermal pressure to the surrounding gas that in the cases of face-on stripped W0 and W45 galaxies acts against the ICM wind. In our simulations, even as the SFR in SFW0 and SFW45 quickly decreases, we might expect the feedback to have a systemic effect on the stripping rate prior to complete quenching (quenching only happens after 500 Myr). The role of feedback is even less clear in the SFW90 galaxy, where the SFR is always 2--3 times higher than in SFNW galaxy, and hence, there is no obvious star-formation-related reason for the long-term fluctuations of the stripping rate ratio. To explain the behaviour we see, we now take a closer look at the gas in their discs.

\begin{figure*}
\centering
\includegraphics[width = 160mm]{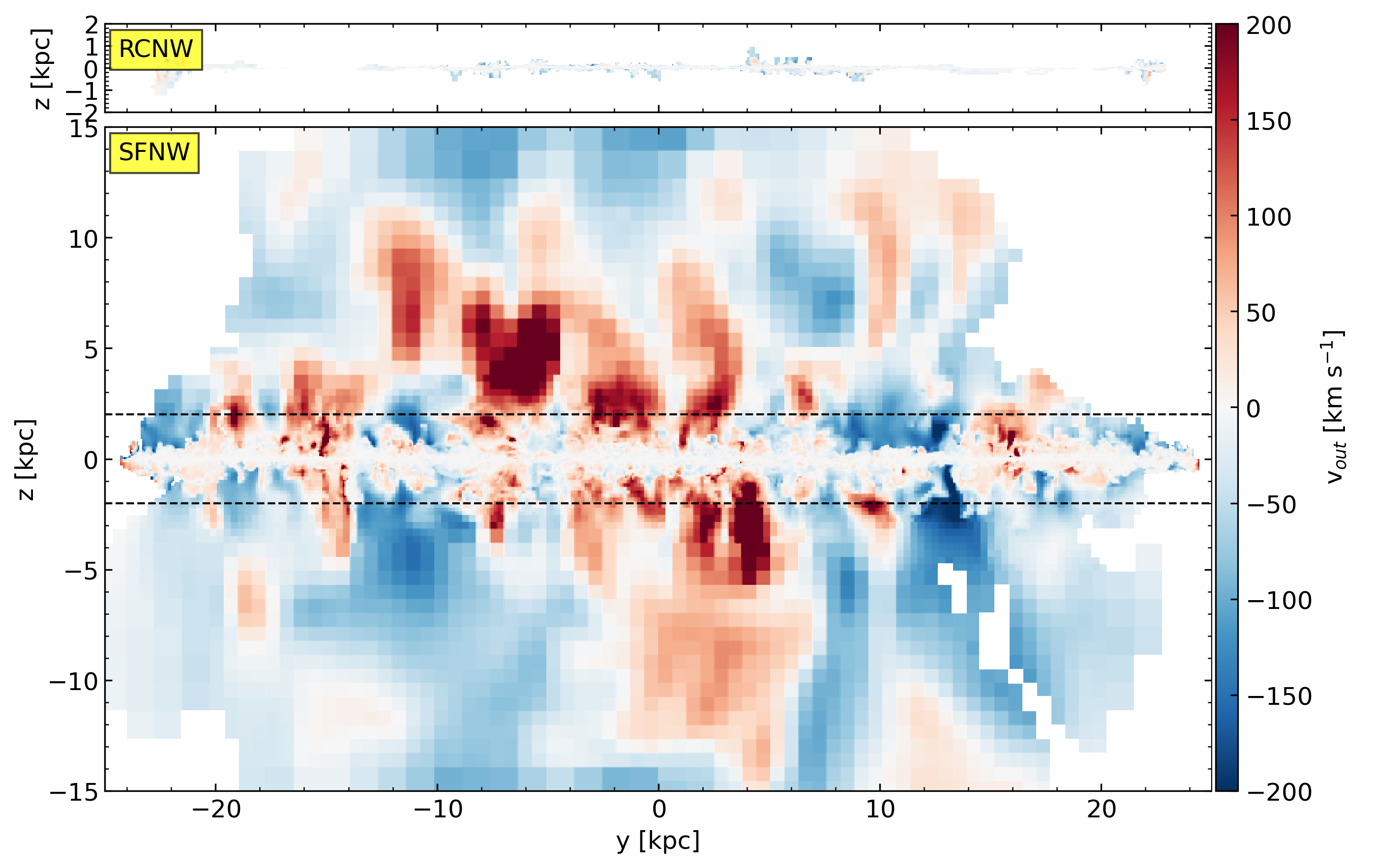}
\caption{Mean cell-mass-weighted outward $z$-velocity, $v_\text{out}$, (outward velocities $v_\text{out} > 0$ and inward velocities $v_\text{out} < 0$) of ISM ($Z>0.35Z_\odot$) in a thin ($x=\pm2$ kpc) slice of the disc for RCNW (top) and SFNW (bottom), taken at 0 Myr. In the bottom panel the black dashed lines indicate the galaxy boundaries, as we define them with $z=\pm2$ kpc.}
\label{fig:NW_vz}
\end{figure*}

\section{What drives the differences between RC and SF galaxies} \label{sec:mechanism}

Here we directly compare the RC and SF simulations in the NW, W0 and W45, and W90 cases separately. We begin by examining the impact of star formation and feedback on an isolated galaxy, build a physical interpretation of how it will affect stripping rates, and then use our understanding to examine the differences in the stripping rates in detail.

\begin{figure}
\centering
\includegraphics[width = 80mm]{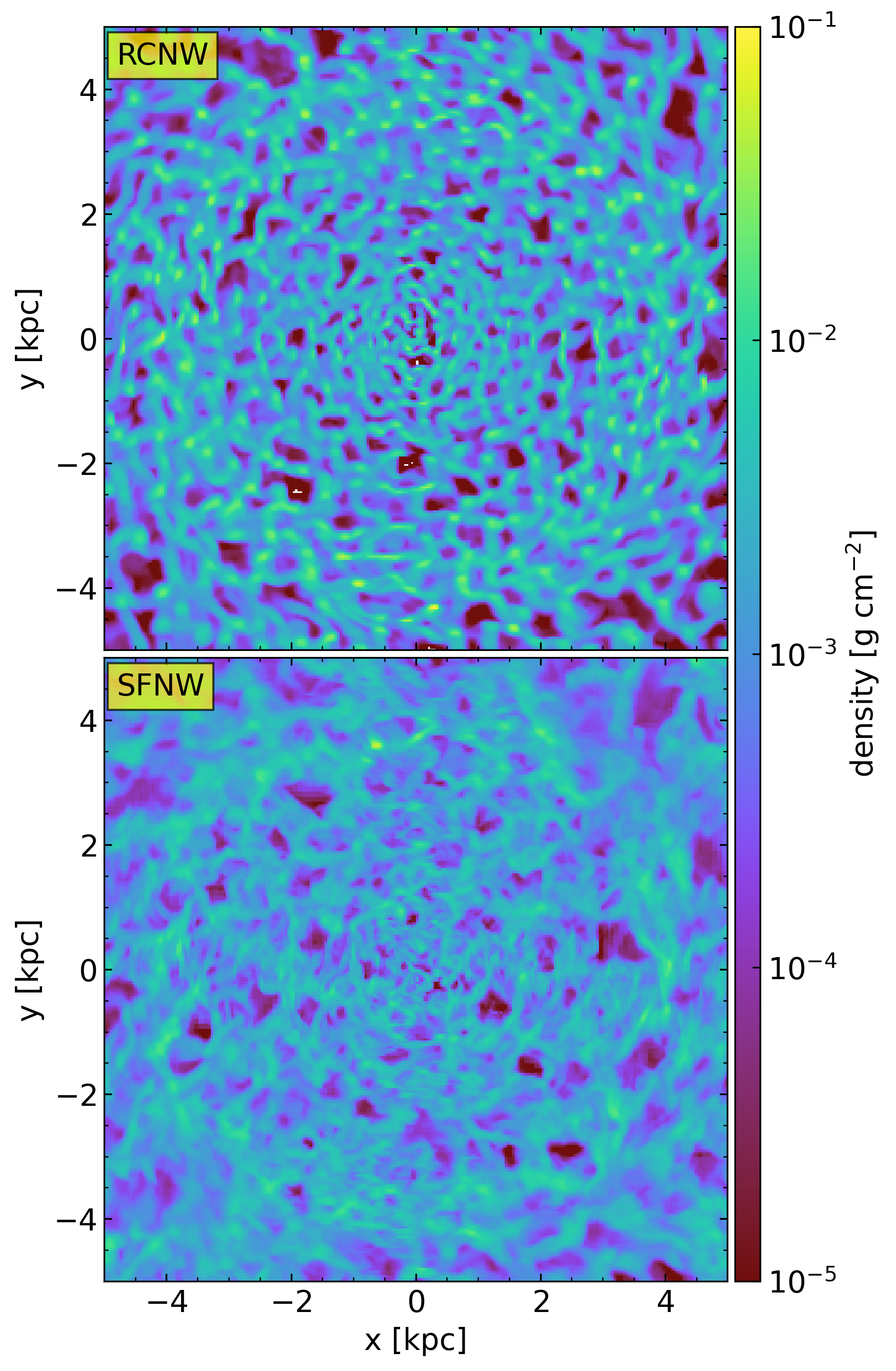}
\caption{Density projections for ISM ($Z > 0.35 Z_\odot$) in the central 5 kpc of the disc, within $\pm2$ kpc of the disc plane for RCNW (top) and SFNW (bottom) at 0 Myr. While star formation and stellar feedback locally smooth out the gas distribution in the disc, in RCNW the gas distribution remains patchy, with holes between the dense clumps.}
\label{fig:NW_dens_proj}
\end{figure}

\subsection{General impact of stellar feedback} \label{subsec:no-wind}

We will start by analysing the impact that star formation and stellar feedback have on gas in general, using our isolated simulations. To illustrate the effect of SN feedback on gas flows around our simulated galaxies, in Fig. \ref{fig:NW_vz} we plot the mean cell-mass weighted outward $z$-velocity, $v_\text{out} \equiv v_z \text{sgn}(z)$ (outward velocities from the galaxy $v_\text{out}>0$ are red and inward velocities to the galaxy $v_\text{out} < 0$ are blue), of ISM ($Z>0.35Z_\odot$) in a thin ($x=\pm2$ kpc) edge-on slice of the RCNW (top) and SFNW (bottom) discs, taken at 0 Myr of RPS (200 Myr total evolution of RCNW and 300 Myr for SFNW). As we begin the simulation, the initially homogeneous ISM starts collapsing into a thin disc, as shown in the top panel of Fig. \ref{fig:NW_vz}, where the disc is about 400 pc thick. While there is no mechanism to stop the collapse into dense clumps in RCNW (illustrated also with the blue inward flows), in SFNW the feedback expels the SN gas from the galaxy via galactic fountains. The bottom panel in Fig. \ref{fig:NW_vz} illustrates these fountains, as they eject the gas (red) and it later falls back on to the galaxy (blue). Along with ionising radiation from young stars and winds from evolved massive stars the fountains act against collapse, and the SFNW galaxy disc becomes much thicker compared to its RCNW counterpart. Although the outflows can eject some material beyond 10 kpc from the disc plane, we highlight that there is very little gas mass in these flows, with only about 0.1 per cent of the total gaseous disc mass extending above 5 kpc from the disc plane.

In addition to this, star formation and stellar feedback also smooth out the gas distribution in the disc, while in the RC galaxies there is no mechanism to inject energy into cold clumps. As a result, the RC gas distribution is patchy, with low-density holes between the dense clumps that would otherwise turn into stars. To illustrate this, in Fig. \ref{fig:NW_dens_proj} we plot the density projection of ISM ($Z > 0.35 Z_\odot$) in the central 5 kpc of the disc within $\pm2$ kpc of the disc plane for the RCNW (top) and SFNW (bottom) at 0 Myr. We stress, however, that the feedback is only able to smooth out the ISM \textit{locally} and does not cause radial mass redistribution across the whole disc.

Another feature present in RC galaxies is a `ring', which manifests itself at $\sim$200 Myr as a non-fragmented density distribution (therefore without high-density clumps) located at $14\text{kpc}<r<20\text{kpc}$. It can be seen in the top panel of Fig. \ref{fig:NW_vz} as a white `gap' where the gas is not moving, and is a result of a non-uniform collapse of the initial disc. The ring is a numerical effect present in many galaxy-scale idealised simulations \citep[e.g.][]{Goldbaum15, Behrendt15} that disappears with time (by 300 Myr of total simulation time or 0 Myr of RPS) in SF simulations as the gas distribution becomes more homogeneous and less clumpy due to SN feedback. In RCNW the ring disappears by 350 Myr of total simulation time, when the whole galaxy disc collapses into dense clumps. Both the added clumpiness in the central disc and the more uniform density distribution in the unfragmented ring play a role in the comparative stripping rates of the RC and SF galaxies.

To summarise, star formation and stellar feedback homogenised the gas distribution in the galaxy disc and prevented it from collapsing into the highly dense clouds found in the inner regions of RCNW, with SN fountains launching the gas as far as 15 kpc above the plane of the galaxy (Fig. \ref{fig:NW_vz}).

\subsection{The Importance of the Disc Gas Distribution}

\begin{figure}
\centering
\includegraphics[width = 85mm]{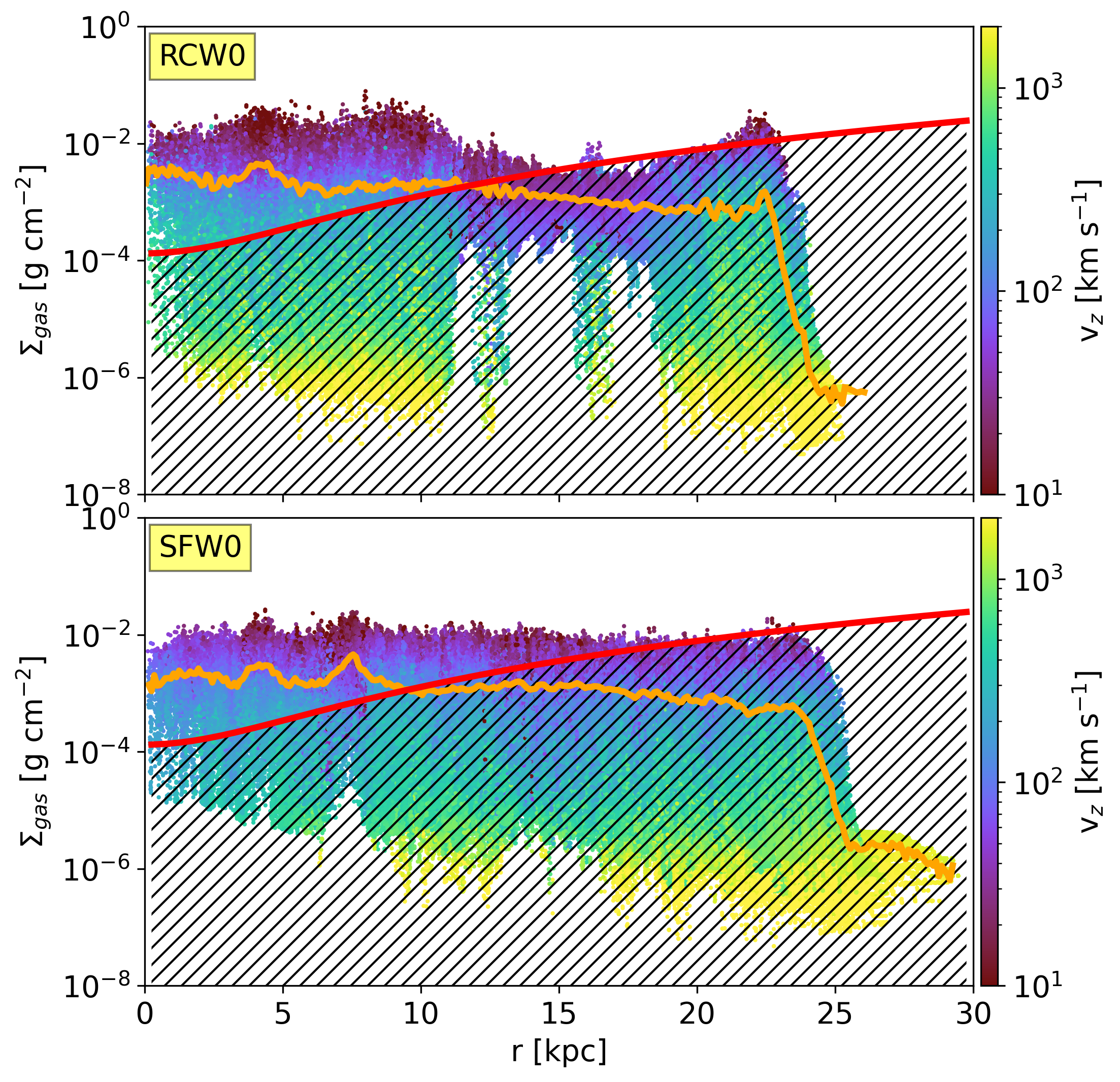}
\caption{Surface density of ISM ($Z > 0.35 Z_\odot$) as a function of galactocentric radius for RCW0 (top) and SFW0 (bottom) at 10 Myr. Surface density is found within our galaxy disc boundaries $z\pm2$ kpc (see the text for details). The points are coloured by the mean cell-mass-weighted $v_z$. The orange line plots average gas surface density as a function of radius. The red line represents threshold surface density from the \citet{GG72} model below which (hatched area) the gas could be stripped at the current RP. This line illustrates that at the galaxy outskirts even the highest density gas will be removed by the ICM wind, while at the centre the gas is protected.}
\label{fig:surf_dens_W0_10}
\end{figure}

To gain a physical understanding of how the gas distribution could affect a galaxy's gas stripping rate, we consider the gas distribution in the disc and the outside-in nature of RPS, which can be quantified by calculating the stripping radius at any time. \cite{GG72} define the stripping radius as the radius at which the gravitational restoring force per unit area equals the RP: $P_\text{ram} = 2\pi G \Sigma_\text{star} \Sigma_\text{gas}$, where $\Sigma_\text{star}$ is the stellar surface density and $\Sigma_\text{gas}$ is the gas surface density. While other equations can be used to calculate whether gas will be removed \citep[e.g.][]{McCarthy08}, this simple criterion has been shown to match reasonably well with simulations \citep{Roediger&Bruggen06, Steinhauser16, Lee20} and is enough to guide our understanding. 

Here, to find the restoring force we measure $\Sigma_\text{gas}$ as a function of radius along the wind direction ($z$-axis), within $z=\pm2$ kpc (the galaxy boundaries). We have repeated the calculations for the disc height of $z\pm5$ kpc and found qualitatively similar results. We measure the $\Sigma_\text{stars}$ only from the static stellar disc, since compared to it the surface density of our star particles is negligible. Assuming instantaneous stripping, this is the radius to which a galaxy would be stripped at any given RP. In our galaxies, immediately after the onset of RP, the stripping radius is $\sim15$ kpc, and due to the increasing RP it drops down to 10 kpc in the next 100 Myr. Note that in this paper we define the stripping radius solely by comparing the gas surface density in our simulated discs to the surface density at which the \cite{GG72} criterion would predict stripping. It is not a measure of where the gas is actually stripped from the disc in our simulation, and indeed the goal of this section is to predict when and understand why simulated galaxies do not follow the simple \cite{GG72} prescription.

The use of the stripping radius to calculate where the ISM will be removed assumes that gas only varies in surface density as a function of radius, which is clearly an oversimplification. Fig. \ref{fig:NW_dens_proj} shows that in both SF and RC galaxies gas fragments into dense clumps embedded in a lower density ISM. To understand the impact of the level of homogeneity in the gas density distribution on gas removal by RPS, in Fig. \ref{fig:surf_dens_W0_10} we plot surface density of the ISM ($Z > 0.35 Z_\odot$) as a function of galactocentric radius for RCW0 (top) and SFW0 (bottom) at 10 Myr. In order to use a more spatially-refined density distribution in the disc, we select the area of $60 \times 60 \text{ kpc}^2$ around the galaxy centre and divide it into $1024 \times 1024$ squares (we confirm that the following results are independent of the resolution). We then integrate the density in each square over the galaxy height of $z\pm2$ kpc. We find qualitatively similar results for the disc height of $z\pm5$ kpc. In contrast with finding an average surface density at a given radius, this method allows us to capture high-density clumps and low-density holes in RCW0 and SFW0, while the choice of surface density instead of the volume/number density informs us about the stripping using the \cite{GG72} criterion. We colour the points by the mean cell-mass-weighted $v_z$. The orange line plots average gas surface density as a function of radius. The red line represents threshold surface density from the \cite{GG72} model. This threshold surface density is found by equating the restoring force to the RP and calculating for each radius the surface density below which (hatched area) the gas could be stripped at the current RP. This line illustrates that at the galaxy outskirts even gas with surface densities well above the average will be removed by the ICM wind, while at the centre nearly all of the gas is protected. 

In this figure we first notice that the surface density in both the RC and SF galaxies varies by orders of magnitude at every radius. Also, although the average value of the surface density is quite similar in both galaxies, the RC galaxy has a larger range of $\Sigma_\text{gas}$ at most radii, except for in the ring region that we introduced in Section \ref{subsec:no-wind}. This figure also shows that SFW0 disc is more radially extended ($r>25$ kpc) than the RCW0 disc with low $\Sigma_\text{gas}$ from stellar feedback. We again emphasise that this is a local effect that does not cause mass redistribution over large radial scales, nor does feedback cause the disappearance of the ring region that happens in RCNW at a similar rate compared to SFNW -- this naturally happens over time due to fragmentation from radiative cooling. We also point out that lower $\Sigma_\text{gas}$ tends to have a higher v$_z$.

We can now discuss how the gas inhomogeneity in the discs will influence gas stripping, starting at the galaxy outskirts. Beyond about 15 kpc, gas must have a higher $\Sigma_\text{gas}$ than the average in order to remain unstripped according to our criterion. In Fig. \ref{fig:surf_dens_W0_10} we see that at the outermost regions of the galaxy ($\sim$23 kpc), RCW0 has more gas at $\Sigma_\text{gas}$ above the red line. Therefore, at early times, when the outermost gas is removed, we would expect SFW0 to be stripped more quickly.

In the ring region of RCW0, at 14--20 kpc, the gas has not yet fragmented and therefore largely lies below the red line. In the same region in SFW0, there is gas at higher $\Sigma_\text{gas}$ that is likely to survive longer. Therefore, in the outer region, we predict that gas homogeneity increases gas stripping as more gas falls below the red line. 

In the inner region we might expect the gas homogeneity to assist RPS as it does in the disc outskirts, but Fig. \ref{fig:surf_dens_W0_10} reveals a different picture. Within the inner 10 kpc gas inhomogeneity leads to some regions of the disc having very low $\Sigma_\text{gas}$. While the mass in these regions is low, the extremely low $\Sigma_\text{gas}$ ISM in RCW0 has high velocity in the wind direction ($v_z$). In fact, in the inner disc, the gas below the red line in RCW0 has a higher $v_z$ than the same $\Sigma_\text{gas}$ ISM in SFW0. Although there are similar amounts of mass below the gas surface density stripping threshold, the higher velocities in the RC ISM result in higher mass fluxes.

We highlight that this is a snapshot of the ISM $\Sigma_\text{gas}$ distribution, and this distribution will vary with time over the simulation. The most obvious differences will be a lack of low $\Sigma_\text{gas}$ ISM at large radii as that gas is removed, and that as RP increases, the red line will shift to higher $\Sigma_\text{gas}$ values. However, all of the general trends we have outlined here persist over time: stellar feedback continues to allow for low $\Sigma_\text{gas}$ gas just beyond the bulk of the disc, dense clumps survive at large radius, and low-density regions are accelerated to high velocities. In the next subsections we will see how this plays out in our simulations. 

\begin{figure*}
\centering
\includegraphics[width = 175mm]{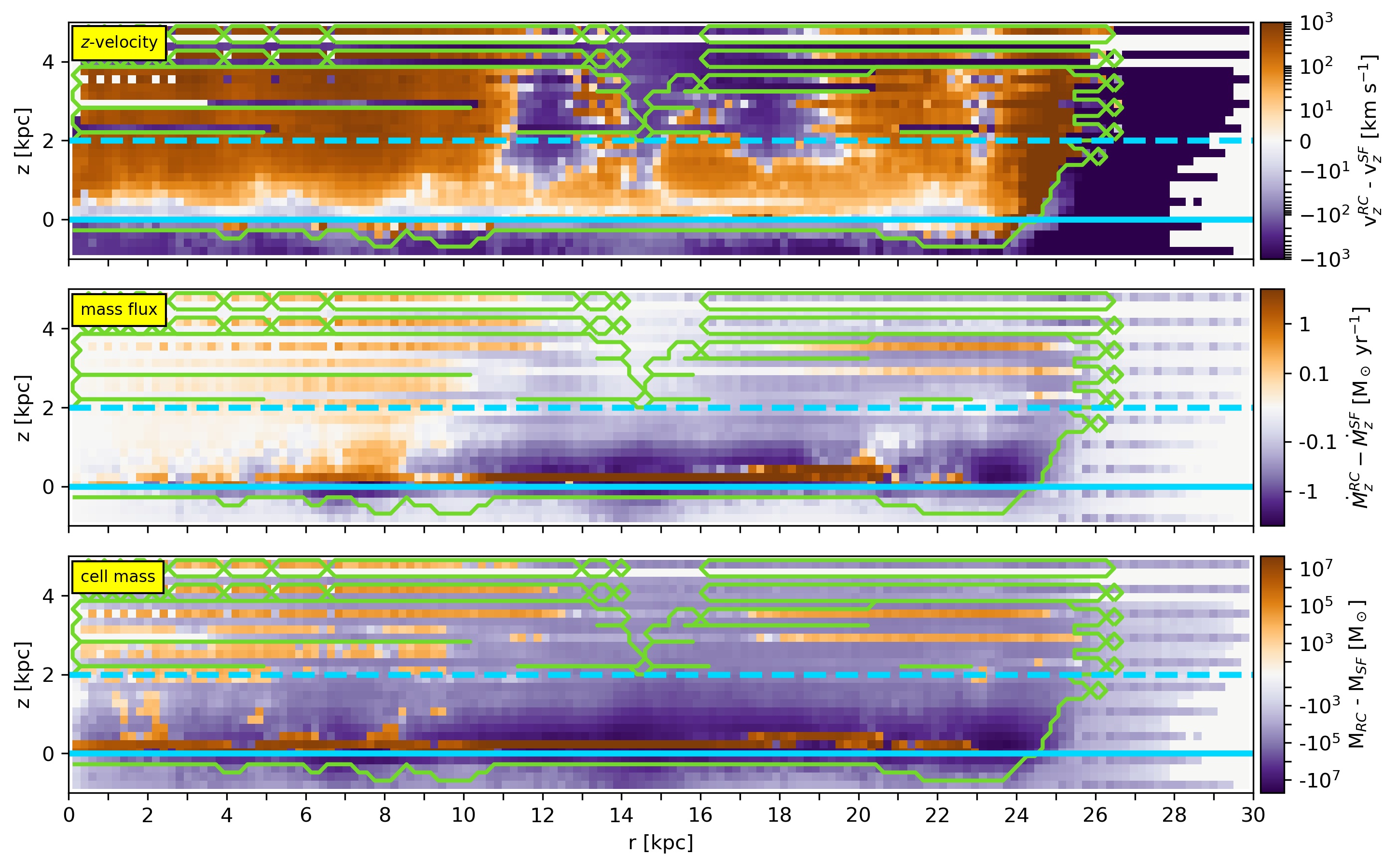}
\caption{For the W0 galaxies at 10 Myr of RPS, from top to bottom: difference between the azimuthally-averaged distributions of cell-mass-weighed velocity $v_z$, vertical mass flux, and total cell mass in RCW0 and SFW0 galaxies, as a function of cylindrical radius $r$ and $z$. We calculate the data only for the stripped gas with $v_z > 0$ to avoid confusion when plotting the difference. The solid cyan line separates downwind and upwind halves of the disc (wind moves upward in the $z$-direction), while the dashed cyan line denotes the the $z=2$ kpc height at which we define the galactic disc boundaries. The green line outlines the gas distribution in RCW0 to illustrate that it is more compact compared to SFW0 at the same time step.}
\label{fig:W0_10Myr}
\end{figure*}

\subsection{Face-on stripped galaxies} \label{subsec:face-on}

Now that we have built intuition for how we expect stripping to be influenced by the ISM gas distribution, we will begin looking at RPS-galaxies, starting with W0, as it is stripped the most rapidly. We measure the mass flux of gas ($Z > 0.35 Z_\odot$) in the galaxy disc ($R=30$ kpc and $h=\pm2$ kpc) in RCW0 and SFW0 galaxies. Then, to understand the evolution of the SF-to-RC stripping rate ratio we will map the difference in the mass flux distribution between the two galaxies. Since in this case RPS is axially-symmetric, we can make use of cylindrical coordinates, focusing only on variations as a function of radius $r$ and $z$-height. Here, we select the data only for the cells with upward outflows $v_z > 0$ to avoid confusion when plotting the difference between the two galaxies (see below), but for completeness, in Appendix \ref{appendix:negative} we show maps for cells with downward inflows $v_z < 0$. The mass flux is:

\begin{equation}
 \dot{M} = \sum_i \frac{m_i v_{z,i}}{dL},
 \label{eq:mass_flux}
\end{equation}

where $m_i$ is mass of the $i$-th cell, $v_{z,i}$ is the $z$ velocity component and $dL=200$ pc is the height of a bin in Fig. \ref{fig:W0_10Myr}, and the sum is extended to all cells in a given $(r,z)$ bin.

We can now spatially compare the mass flux between the RC and SF galaxies. In Fig. \ref{fig:W0_10Myr} we plot for the two W0 galaxies at 10 Myr from top to bottom: difference between the $(r,z)$ maps of cell-mass weighted azimuthally-averaged velocity $v_z$, vertical mass flux, and total cell mass in RCW0 and SFW0 galaxies, as a function of cylindrical radius $r$ and cylindrical $z$. The bin size in each of the panels is $200 \times 200 \text{ pc}^2$. The goal of this plot is to map which of the two galaxies, SFW0 or RCW0, has more mass flux in the wind direction (central panel), and which of the mass flux components contributes more to it. Each panel shows RC $-$ SF, so orange indicates higher RC values, and purple indicates higher SF values. With the solid cyan line ($z=0$) we separate downwind and upwind halves of the disc, while the dashed cyan line denotes the $z=2$ kpc height at which we define the galactic disc boundaries. We also overplot a green contour that outlines the gas distribution in RCW0 as it would appear if in this figure we plotted real RCW0 values\footnote{We note that these green contours are only denoting the edge of the distribution of disc gas, defined as gas with a metallicity above 0.35. There is gas in the cells beyond this contour, but it is ICM gas.} instead of the difference. This illustrates that RCW0 is more compact compared to SFW0 at the same time step, since all gas outside the contour is only from the SFW0 disc.

As mentioned earlier, the SF-to-RC stripping rate ratio for W0 galaxies changes periodically in the first 200 Myr. Here we look in detail at these different periods and what causes the change, always referring to the top panel of Fig. \ref{fig:strip_mass}. At the beginning, up to 25 Myr, SFW0 is stripped faster, illustrated in the middle panel of Fig. \ref{fig:W0_10Myr} with purple outflows beyond $r>10$ kpc. At $r<10$ kpc, however, the difference between the RC and SF mass fluxes is quite small but tends towards higher mass flux in the RC simulation. Interestingly, the inner region of the SF galaxy has stronger outflows from feedback (Fig. \ref{fig:NW_vz}), so those feedback-driven outflows do not seem to be dominating the mass flux. Close to the disc plane we see enhanced mass flux in the RC galaxy, which will result in larger RC stripping rates at later times.  

Since stellar feedback expels gas in all directions, the gas disc of SFW0 is much thicker and more radially extended (by $\sim2$ kpc) than in RCW0, as shown in Fig. \ref{fig:NW_vz} and in the bottom panel of Fig. \ref{fig:W0_10Myr}. This added thickness does not change the gas surface density, and therefore does not affect the stripping rate. The feedback does not directly remove gas or assist RP in doing so, as it would be with the gas getting an additional `kick' to its vertical velocity by galactic fountains, but, rather, enhanced stripping is a result of star formation and stellar feedback homogenising the ISM (as illustrated in Fig. \ref{fig:NW_vz} and \ref{fig:NW_dens_proj}), and radially extending the gas disc (as seen in Fig. \ref{fig:surf_dens_W0_10}). We note that the homogeneity and the added radial extent persist in the SF runs because star formation and feedback continually affect the ISM.

Although most of the gas is lost from the outskirts and the gas disc generally shrinks as the RP continuously increases, these figures can also be used to explain the difference in stripping rates in the inner disc. In the inner 10 kpc the SFW0 gas seems to be protected from RPS, as the mass-flux distribution is dominated by the high-velocity gas from RCW0 (orange area in the top panel of Fig. \ref{fig:W0_10Myr}). The ICM wind can easily flow through the underdense regions created by continuous cooling and gas fragmentation (Fig. \ref{fig:NW_dens_proj} and \ref{fig:surf_dens_W0_10}). As it does so, it rapidly accelerates the low density ISM, and it is this gas that can be stripped even from the galaxy centre that we see in orange in the middle panel of Fig. \ref{fig:W0_10Myr}. 

\begin{figure*}
\centering
\includegraphics[width = 175mm]{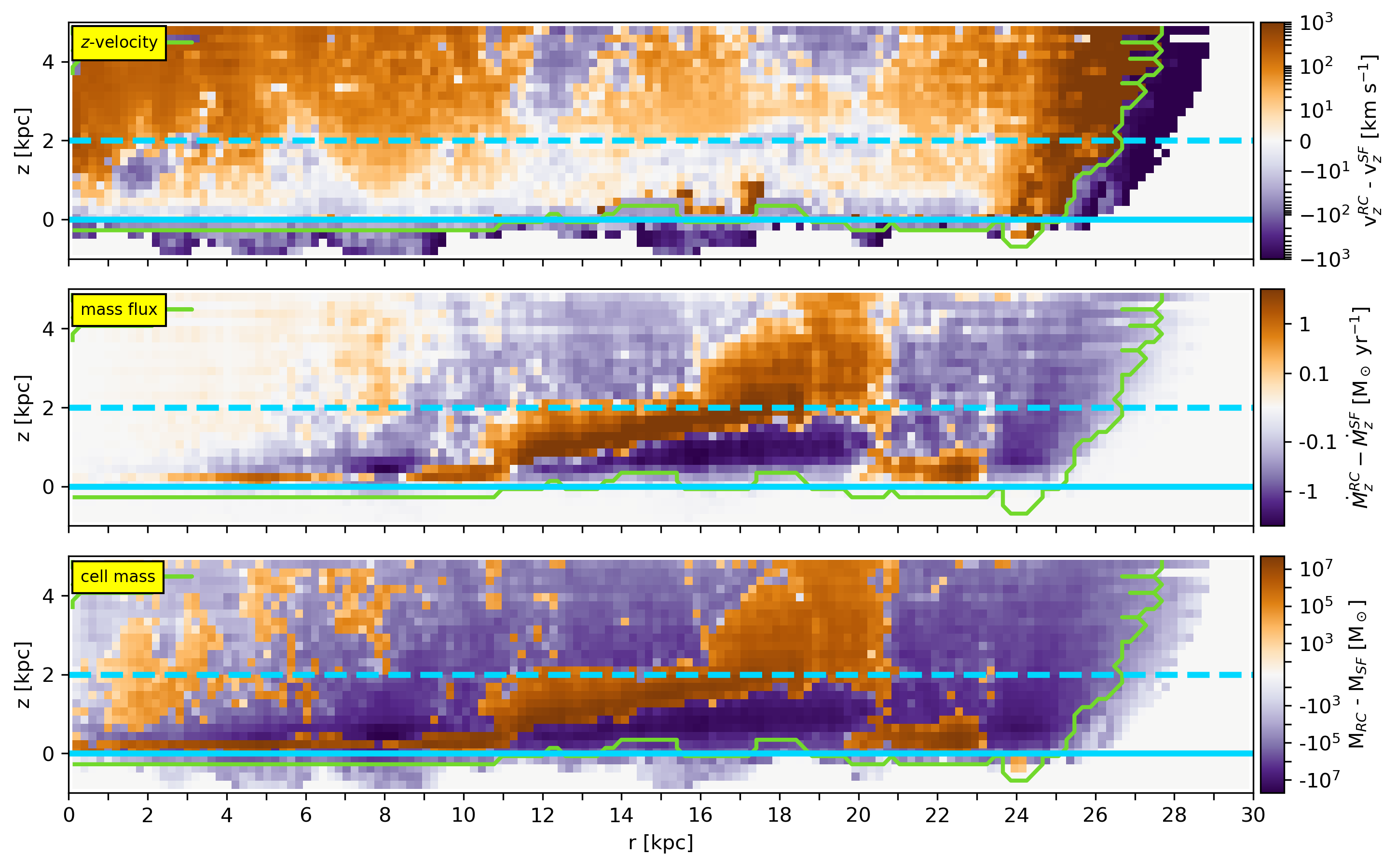}
\caption{Same as Fig. \ref{fig:W0_10Myr} but for 35 Myr of RPS. At this time, the stripping rate is dominated by gas removed from a more uniform, unfragmented low-density region between $14\text{kpc}<r<20\text{kpc}$ in RCW0 galaxy.}
\label{fig:W0_35Myr}
\end{figure*}

In addition to the enhanced mass flux from the SFW0 galaxy crossing $z = 2$ kpc, at 10 Myr we can see enhanced mass flux formed in the RCW0 galaxy in a narrow strip directly above the $z = 0$ line (Fig. \ref{fig:W0_10Myr}, middle and bottom panels, in orange). By 35 Myr, this stripped slice evolves into a whole arch as shown in Fig. \ref{fig:W0_35Myr}. This arch is the reason why the stripping rates are higher in RCW0 during 30--50 Myr and why at that time there is a peak in the top panel of Fig. \ref{fig:strip_rate_sf}. This peak is also present in SFW0 but is due to the gas removal at large radii. The enhanced stripping of the RCW0 ring once again illustrates that, as in the outskirts of SFW0, homogeneity of gas leads to more gas below the stripping surface density (as there is an absence of dense clumps that are present in this region in SFW0), which makes it easier to be stripped.

In the middle panel of Fig. \ref{fig:W0_35Myr} at $12\text{ kpc}<r<20\text{ kpc}$ we can see SFW0 gas that begins to move (the dark purple area under the orange arch). Later, during 55--85 Myr, as SFW0 is stripped outside-in, the same gas is leaving the galaxy. Since by this time there is no gas left in RCW0 at these radii (the gas arch seen here has passed beyond $z=2$ kpc), in this period the stripping rate is stronger in the SF galaxy.

The final change to the SF-to-RC stripping rate ratio happens at 90 Myr. With a smooth ISM distribution, a galaxy subject to face-on RPS will be gradually reduced in size up to its stripping radius. This largely occurs in SFW0, which has a constant peak gas surface density at all radii as shown in Fig. \ref{fig:surf_dens_W0_10}. Therefore, increasing RP smoothly strips it from the outside-in. In RCW0, however, while the loosely-bound gas was easily stripped from the former ring area, the dense clumps of gas in the outskirts at $r>20$ kpc remain until 90 Myr. We can see these dense clouds surviving at 35 Myr in the bottom panel of Fig. \ref{fig:W0_35Myr} by the orange region of excess mass and in Fig. \ref{fig:surf_dens_W0_10} as the high surface density gas at $\sim$23 kpc. Hence, from 90 to 170 Myr while SFW0 has already been reduced to 12 kpc radius, RCW0 once again dominates the stripping rate as the remaining gas clumps in the outskirts are being quickly stripped in addition to the gas at $\sim$12 kpc that is being stripped from both galaxies.

During 170--400 Myr we see similar stripping rates in the RC and SF galaxies. By this point both the W0 galaxies have been reduced to the same stripping radius (using the \cite{GG72} gas surface density definition), around 12 kpc, with the gas distribution in SFW0 still being 1--2 kpc more extended than in RCW0. Because feedback continues to push some of the ISM to larger radii, the SFW0 galaxy is stripped faster in the outskirts. On the other hand, in the galaxy centre RCW0 is stripped faster due to the low surface density regions in the gas distribution, making the overall picture of RPS similar to what we saw during the first 30 Myr. This also agrees well with our intuition derived from the gas surface density distributions seen in Fig. \ref{fig:surf_dens_W0_10}. Therefore, feedback both enhances outer stripping by extending the ISM and depresses inner stripping by homogenising the gas surface density. At this time in our simulations these two effects balance each other out, and globally the galaxies have almost identical stripping rates (Fig. \ref{fig:strip_mass}, top panel).

From 400 Myr onward the SFW0 galaxy starts losing gas more and more slowly. During this time its calculated stripping radius and measured gas disc radius are both smaller than that of RCW0. As shown for earlier times by Fig. \ref{fig:W0_10Myr} and \ref{fig:W0_35Myr} the mass flux is determined mostly by the mass distribution: where there is more mass, the mass flux is stronger, even if that gas is moving slower. By 400 Myr, SFW0 has lost gas not only to RPS, but also to star formation, which is especially true in the inner 5 kpc, where most of the stars are formed. This means that in the later part of the simulation there is simply very little gas left in the centre of SFW0, and therefore RCW0 starts being stripped faster.

As evident from the stripping rates in Fig. \ref{fig:strip_rate_sf} and, consequently, from the mass of the stripped gas in the bottom panel of Fig. \ref{fig:strip_mass}, the W45 galaxies follow a very similar evolution to the W0; something that was also shown by \cite{Roediger&Bruggen06} and \cite{Akerman23}. It is not surprising then that the SF-to-RC stripping rate ratio of W45 galaxies also follows the one for the W0 (Fig. \ref{fig:strip_mass}, top panel). The main difference between the two ratios is due to the time delay that, as explained above, stems from the fact that it takes an inclined wind more time to reach a galaxy and begin the stripping process. Another difference shown in the top panel of Fig. \ref{fig:strip_mass} is that during two relatively short periods of time, 225--250 Myr and 305--375 Myr, SFW45 is stripped faster than RCW45. This is another effect of the continuous collapse of gas, as a result of which dense clumps form. These clumps are then pushed in the shadow of the disc, where they are protected from the ICM wind. The stripping rate in RCW45 drops and the SF-to-RC stripping rate ratio goes up.

In conclusion, the added inhomogeneities in RC galaxies result in more gas both above and below the stripping surface density (Fig. \ref{fig:surf_dens_W0_10}). Dense gas is harder to strip, as evidenced by the lower $v_z$ at high surface density. In the galaxy outskirts, ISM inhomogeneities are able to enhance gas surface density to the point where it cannot be stripped, slowing down the global stripping rate. We see this in the delayed stripping of dense gas beyond 20 kpc in the RC galaxies relative to SF galaxies, as well as in the earlier stripping of the homogeneous gas ring at 14--20 kpc in the RC galaxies. In the galaxy centre, ISM inhomogeneities can result in regions with very low gas surface density. These regions are stripped very quickly, which leads to faster stripping rates in the central regions of RC galaxies.  

\begin{figure*}
\centering
\includegraphics[width = 175mm]{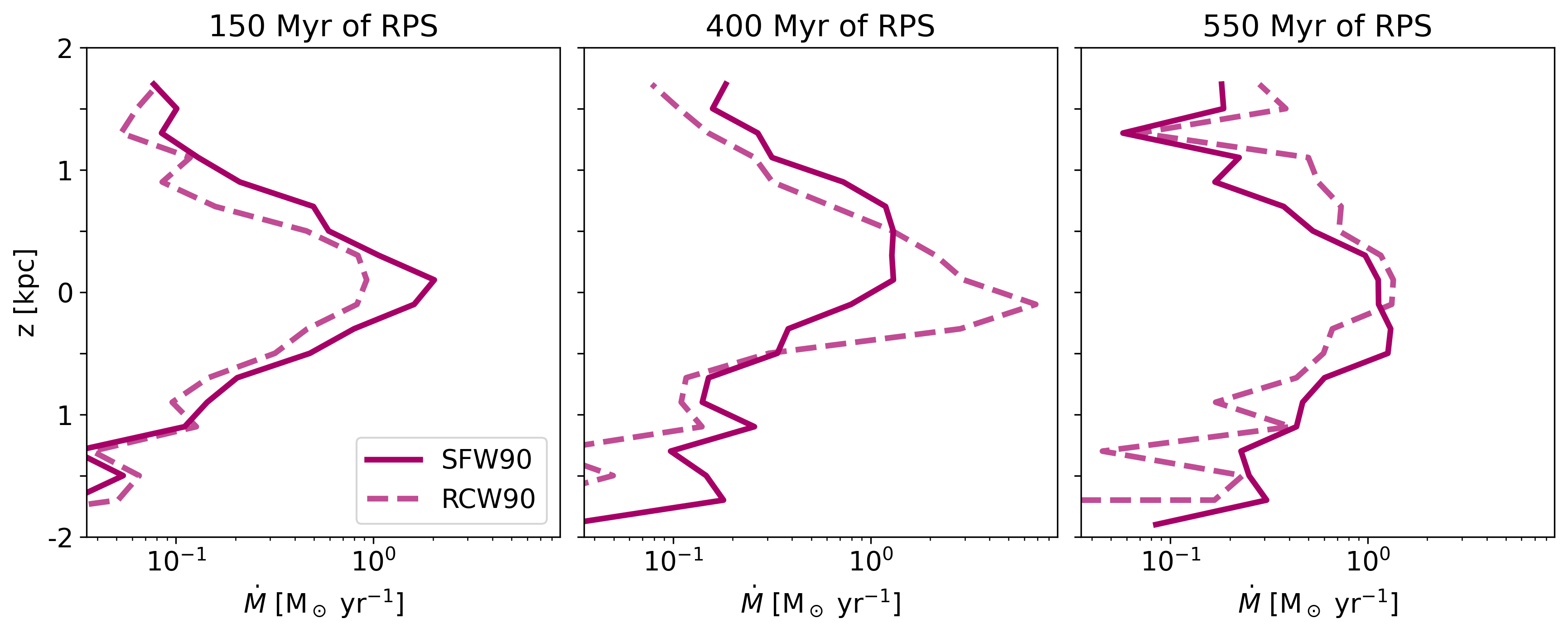}
\caption{For the W90 galaxies, $v_y$ mass flux of ISM ($Z > 0.35 Z_\odot$) as a function of $z$ measured in $29.5 \text{kpc}<r<30.5 \text{kpc}$ at (from left to right) 150, 400 and 550 Myr of RPS for SF (solid) and RC (dashed) simulations. We select the data for the downwind part, $y>0$, and for the stripped gas with $v_y > 0$.}
\label{fig:W90_mass_flux}
\end{figure*}

In this section we have focused on the early times when the RC and SF stripping rates differ in order to understand the physical reason for the distinct rates. However, generally, the lower stripping rates at the outskirts of the RC galaxies balance the higher stripping rates in their centres, leading to similar stripping rates in the SF and RC simulations.

\subsection{Edge-on stripped galaxies}

After having discussed the detailed evolution of the face-on stripped galaxies, we examine the edge-on ones, RCW90 and SFW90. Since in this case the feedback does not act parallel to the direction of the wind, its effect is harder to predict. Still, we could expect RP from the edge-on ICM wind to easily remove the gas expelled above (and below) the disc by the galaxy fountains. Surprisingly, while this effect indeed takes place, it does not dominate the stripping rates in SFW90.

To illustrate this, in Fig. \ref{fig:W90_mass_flux} we show the $v_y$ mass flux (similar to eq. \ref{eq:mass_flux}) measured in $29.5 \text{kpc}<r<30.5 \text{kpc}$ as a function of $z$. We remind the reader that in the edge-on case (W90) the wind moves along the $y$-direction. Solid (dashed) lines plot SFW90 (RCW90) at three different times (from left to right): 150 Myr, 400 Myr, and 550 Myr. We chose these time steps to have three different SF-to-RC stripping rate ratios in the top panel of Fig. \ref{fig:strip_mass}, where at 150 Myr SFW90 is stripped faster, at 400 Myr RCW90 dominates and at 550 Myr the galaxies lose their gas at a similar rate. We also select the data only for the downwind part, $y>0$, and for $v_y > 0$ only. Thus, shown here is that while (for the most part) outside of the disc plane there is more mass flux in SFW90 than in RCW90, the bulk of the stripping at all times is in the galaxy plane.

Moreover, shown in the bottom panel of Fig. \ref{fig:strip_rate_sf} is SFR of SFW90, enhanced compared to SFNW on the whole duration of the simulation. This means that the scenario outlined above of SN outflows being swept away by the ICM wind plays out during all 700 Myr. Globally, evolution of the SF-to-RC stripping rate ratio (Fig. \ref{fig:strip_mass}, top panel) can be divided into two periods: from 120 to 400 Myr the ratio goes down, indicating a slow but steady decrease in the relative outflow rates of SFW90 compared to RCW90; from 400 to 700 Myr it goes up again, meaning that now the SF galaxy gets stripped systematically faster. This means that despite the ongoing stripping of the expelled feedback gas, RCW90 manages to overcome the effect, and is even stripped faster than SFW90 from 330 to 530 Myr (as also evident in the middle panel of Fig. \ref{fig:W90_mass_flux}).

\begin{figure*}
\centering
\includegraphics[width = 160mm]{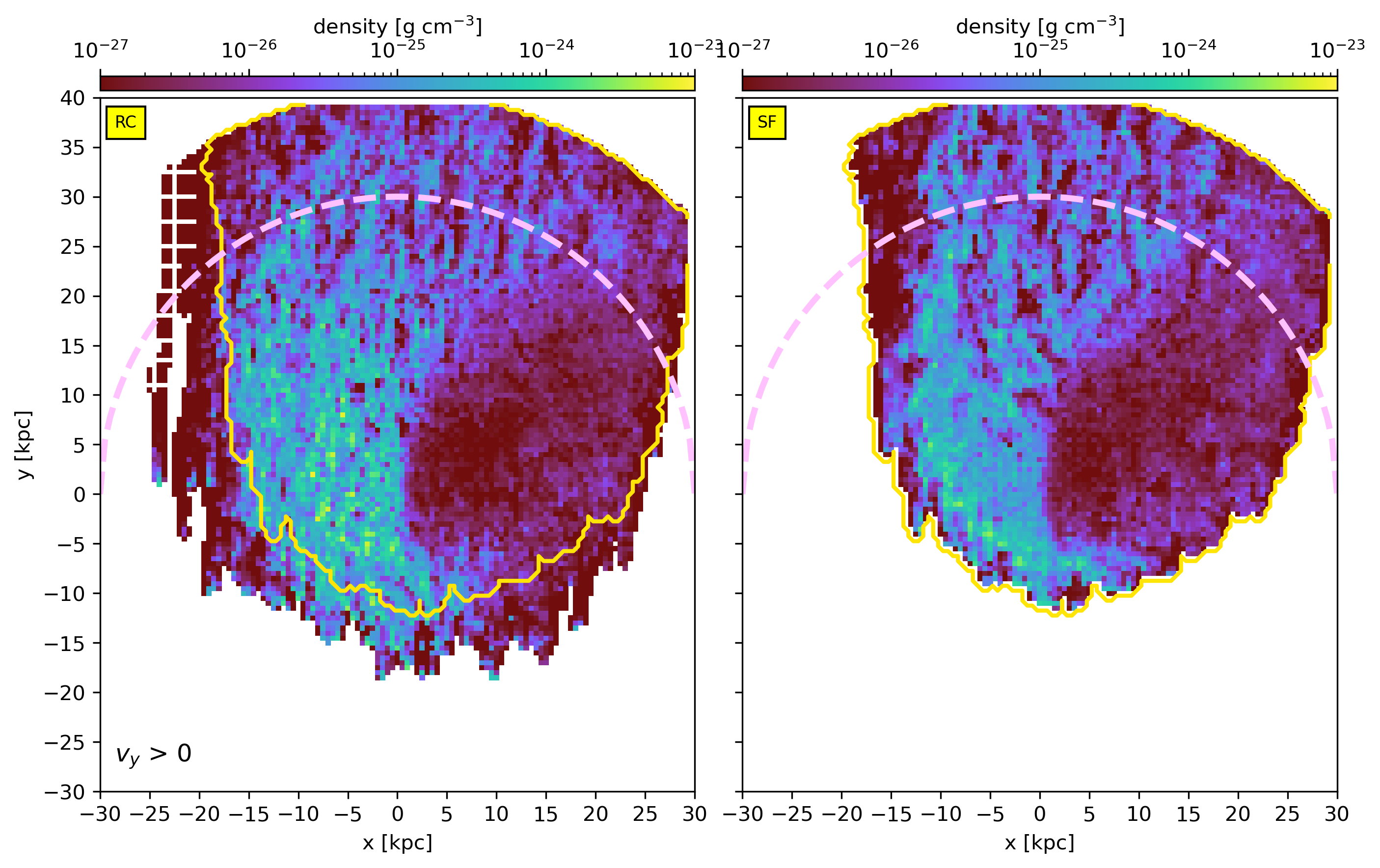}
\caption{Density (cell-volume weighted) distribution of gas ($z\pm2$ kpc) in RCW90 (left) and SFW90 (right) at 250 Myr. We calculate the data only for the stripped gas with $v_y > 0$. The dashed pink line denotes the upper boundary of $r=30$ kpc with which we define the galactic disc. The yellow line outlines the gas distribution in SFW90 to illustrate that it is more compact compared to RCW90 at the same time step.}
\label{fig:W90_250Myr}
\end{figure*}

To understand the change in the stripping rate, we look at the gas ($Z > 0.35 Z_\odot$) distribution in the disc plane ($z\pm2$ kpc), shown in Fig. \ref{fig:W90_250Myr} by an $xy$ density projection for 250 Myr. Again, we calculate the data only for the gas with $v_y > 0$. This leads to a lack of gas at positive $x$-values in the figure, where galaxy rotation leads to negative $v_y$. In both panels, the dashed pink line denotes the outer boundary of $r=30$ kpc with which we define the galactic disc. The yellow line outlines the gas distribution in SFW90 to illustrate that it is more compact compared to RCW90 at the same time step. Also note that unlike in the W0 runs, where the gas needs to travel up to $z = 2$ kpc to be considered stripped, in W90 this path is increased to $r = 30$ kpc. Any change in the gas distribution in the galaxy disc will affect the stripping rate only about 100 Myr later, since the gas needs to travel 30 kpc on average to the downwind edge. This needs to be kept in mind when analysing the plots, since differences in the immediate gas distribution of RCW90 and SFW90 may not reflect the current SF-to-RC stripping rate ratio.

The gas distribution in RCW90 and SFW90 reveals that, just like in the face-on stripped galaxies, the main factor contributing to the differences between SF and RC galaxies is not direct removal of gas from the disc plane due to feedback (such as galaxy fountains) but the homogenised gas distribution in the disc. When hit by an edge-on ICM wind, the dense clouds are hard to remove; at the same time, they shield the easily-stripped low-density gas. Compared to this, the more homogeneous gas in SFW90 is being easily pushed as evident in the top panel of Fig. \ref{fig:strip_mass}. This is reflected in the more extended distribution of gas in RCW90 compared to SFW90 -- the higher density clouds in RCW90 have been less affected by the ICM wind and therefore less gas has been removed. 

However, while the lower density ISM in SFW90 has been stripped more quickly than the higher density clouds in RCW90, the ISM is not simply directly stripped from the galaxy. As described above, the gas has a long way to travel before being considered `lost' by the galaxy (the longest path from the upwind to the downwind side is $\sim55$ kpc). On this course, the combination of wind and galaxy rotation makes gas from different galactic radii collide with each other to create a dense `arch' (right panel of Fig. \ref{fig:W90_250Myr}, at the leading edge of the disc in the bottom left quadrant). At this point, the SFW90 galaxy is also substantially smaller than RCW90, as outlined by the yellow line. As the arch forms, its dense gas becomes harder and harder to strip, at which point the SF-to-RC stripping rate ratio crosses the equality line. In the bottom panel of Fig. \ref{fig:strip_rate_sf} the decreasing $\dot{M}$ after 330 Myr indicates that SFW90 runs out of easily stripped gas at this point. The remaining gas from the arch is pushed into a big dense knot on the wind-facing side of the disc about 10 kpc from the galaxy centre. At this point the stripping rate in SFW90 drops, while the RC galaxy maintains its stripping rate by removing the low-density gas.

The steady pushing of gas to the galaxy centre also proceeds in RCW90, but more slowly since the dense clumps can only be pushed by higher RP (which occurs later in the simulations). By 400 Myr RCW90 repeats the fate of SFW90: gas forms an even denser (compared to the SF simulation) knot in the galaxy centre, and after that the stripping rate starts to drop (Fig. \ref{fig:strip_rate_sf}, bottom panel).

In the next $\sim150$ Myr, SFW90 slowly overtakes RCW90 in the stripping rate (Fig. \ref{fig:strip_mass}, top panel). The dense central knot will inevitably be destroyed by the increasing RP (as it starts to at $\sim$550 Myr), and its gas stripped. From $\sim550$ Myr onward the two galaxies are stripped at about the same rate.

In conclusion, the main differences between the stripping process in the edge-on stripped galaxies, just like in the face-on ones, stem from the fact that in the SF galaxies the gas is homogenised and thus more easy to push at first, while in the RC galaxies it is patchy with dense clumps that are comparatively harder to move. This however does not simply mean that SFW90 is always stripped faster, because the edge-on RP can compress gas to higher densities as well as directly remove gas, and once the high density region in the SF galaxy matches that in the RC galaxy, stripping occurs at the same rate.

\section{Discussion}\label{sec:discussion}

Our simulations show that the stripping rates of star-forming and radiative-cooling-only galaxies do not differ significantly, and that on galactic scales the stellar feedback does not directly assist or impede RPS by driving significant gas mass outside the disc plane (Fig. \ref{fig:strip_rate_sf}). While there are not many papers that study the effect of stellar feedback on the process of stripping, the ones that do, \cite{Bahe&McCarthy15} and \cite{Kulier23}, show results different from ours. Both of these works explore cosmological smoothed particle hydrodynamic simulations, the former with Galaxies-Intergalactic Medium Interaction Calculations (GIMIC) and the latter with EAGLE data.

\cite{Bahe&McCarthy15} reveal two things. Firstly, the fraction of gas that can be stripped from a galaxy declines with galaxy mass, while the fraction that can be removed (but still remain bound to the galaxy halo) via galaxy fountains, on the contrary, increases. If we consider their galaxies that are the most similar to ours, with stellar masses $M_\text{star} \approx 10^{10.75} M_\odot$ in massive clusters with $M_\text{cluster} = 10^{15.2} M_\odot$, then it appears that RPS accounts for 60 per cent of all removed gas, while the rest is due to galactic fountains. Secondly, they also show that the stellar feedback implemented in GIMIC (and EAGLE) gave a prior `kick' to a significant part of the stripped gas that later allowed RPS to quickly remove it. Still, for the most massive galaxies in the most massive clusters the amount of stripped gas when accounting for the `kick' is within the boundaries predicted by \cite{GG72} model. \cite{Kulier23} come to a similar conclusion, where they measure the `excess' gas that is stripped due to the feedback and reveal that the amount increases with RP. This indicates a synergy between RP and stellar feedback in removing galactic gas.

These findings are, perhaps, similar to our own, but rather than the direct effect of SN feedback, SFW0 is stripped faster at the galaxy outskirts due to the feedback homogenising and radially extending gas in the galaxy disc, thus lowering its maximum surface density. However, we also show that in a direct comparison with a non-star-forming galaxy this effect proves to be insignificant. It is worth emphasising again that both \cite{Bahe&McCarthy15} and \cite{Kulier23} use cosmological smoothed-particle hydrodynamics simulations. By their nature such simulations cannot have high resolution and sophisticated feedback physics of galaxy-scale adaptive-mesh-refinement simulations. Due to these necessary technical limitations the effect of stellar feedback might be overestimated. For example, in EAGLE the feedback is ejective, with galactic fountains launching gas to large distances \citep{Mitchell20}; it is also known to create excessively large `holes' in the ISM \citep{Bahe16}. In this light, the role of the homogenised ISM density distribution in RPS is likely underestimated in these simulations.

\cite{Bahe&McCarthy15} may give us insight into the possible galaxy mass dependence of our result. They find that while the highest-mass galaxies seem to be the least affected by the stellar feedback when it comes to stripping, the lower-mass galaxies $M_\text{star} < 10^{10} M_\odot$ lose much more gas than estimated by \cite{GG72}. \cite{Muratov15} in a cosmological zoom-in simulation FIRE measured the mass loading factor (mass carried out by stellar feedback per mass of newly formed stars) and found a decreasing trend with galaxy stellar mass, although their galaxies do not reach the stellar mass of our simulated galaxy. This means that low-mass galaxies lose more gas via galaxy fountains, in a seeming agreement with \cite{Bahe&McCarthy15}. In this case, it is possible that stellar feedback would assist the gas removal via RPS of low-mass galaxies. We stress that our specific result that SN feedback has little effect on the stripping rate of galaxies may be quite mass dependent. However, our result that the homogeneity of the ISM will have important consequences on the radial distribution of stripping is general. 

Another important note about this work is that we did not test a range of feedback models, in which we would expect feedback implementations that strongly eject mass to large distances from the disc to have a bigger effect on the stripping rate of galaxies. However, we can compare our results with other works studying SN feedback. \cite{Kim20} conducted local simulations of several regions in a Milky-Way-like galaxy with a resolution of 2 to 8 pc. Similar to \cite{LiBryan20} they find that the hot ($T>10^6$ K) galactic outflows are much more energetic than the cool ($T<10^4$ K) ones, which, conversely, are responsible for driving the mass out of a galaxy. We can compare mass flux from different regions in SFNW that correspond to the regions in \cite{Kim20}. We find that at the radius where our $\Sigma_\text{gas}$ and $\Sigma_\text{star}$ are similar to those in \cite{Kim20} (R8), our mass fluxes are also comparable to theirs. This therefore gives us confidence that the scale of our mass fluxes is reasonable, and our galaxy should not have dramatically different mass fluxes that would change our main results.

We also reiterate that these are purely hydrodynamical simulations, so we do not include magnetic fields or cosmic rays. Magnetic fields have been shown in simulations to broaden the density PDF \citep{Burkhart15}, especially for low-Mach number turbulence \citep{Burkhart09}, and may decrease the SFR \citep{FederrathKlessen12, Federrath15, Wibking23}. There is also some observational evidence that enhanced magnetic fields may reduce gas fragmentation \citep[e.g.][]{Anez-Lopez20}. Simulations also indicate that cosmic rays smooth the distribution of gas density in the ISM and the circumgalactic medium \citep{Butsky&Quinn18, Butsky20, Farber22}. The impact of these sources of non-thermal pressure on RPS could be an interesting avenue for future work.

\section{Conclusions} \label{sec:conclusions}

In this paper we present a suite of wind-tunnel galaxy-scale simulations to determine the role of star formation and stellar feedback in stripping a galaxy of its gas via RP. The initial conditions are the same in each simulation, and model a galaxy slightly more massive than the Milky Way ($M_\text{star}=10^{11}M_\odot$) falling into a massive cluster ($M_\text{cluster} = 10^{15} M_\odot$). We have two sets of simulations: radiative-cooling-only (denoted as RC) and with star formation and stellar feedback (SF). In each of the two sets there are four galaxies, of which one is not undergoing RPS (no wind, NW). The other three galaxies are on their first infall into a cluster and have different galaxy-wind impact angle: face-on (W0), edge-on (W90) and $45^\circ$ (W45). We compare the stripping histories of a RC and SF galaxy at the same impact angle and find that:

\begin{enumerate}

\item Star formation and stellar feedback vertically puff up the galaxy disc, expelling galactic fountains up to 15 kpc, and homogenise the ISM density distribution. In RC galaxies, the gas continuously cools down and fragments into dense clumps (Fig. \ref{fig:NW_vz} and \ref{fig:NW_dens_proj}), resulting in a broader distribution of gas densities in the ISM. In addition, since the fountains eject gas in all directions, including radially, at 0 Myr (the time when the stripping begins) the radius of a SF galaxy is $\sim2$ kpc bigger than that of a RC one.

\item Globally, the stripping rates are quite similar for RC and SF galaxies of the same impact angle (Fig. \ref{fig:strip_rate_sf}), although since part of the gas is used up for star formation, the SF galaxies lose less total gas through RPS (Fig. \ref{fig:strip_mass}, bottom panel). Comparing how the stripping rates change relative to each other (Fig. \ref{fig:strip_mass}, top panel) reveals that the relative stripping rates of the SF and RC galaxies invert multiple times over the course of the simulations.

\item The differences in the stripping rates of the RC and SF galaxies are related to the differences in the density distributions in the galaxy discs. This is because a large distribution in gas density results in a large distribution in gas surface density (Fig. \ref{fig:surf_dens_W0_10}). While the average surface density is not affected, more inhomogeniety in the ISM of RC galaxies results in more gas both above and below the stripping surface density at which the restoring force equals the RP. Gas that falls below the stripping surface density may be more easily removed, while the gas above it may survive unstripped.

\item In the outskirts of W0 and W45 where the average surface density is below the stripping surface density, inhomogeneity makes stripping slower by allowing dense clouds to survive (Fig. \ref{fig:surf_dens_W0_10} and \ref{fig:W0_10Myr}).

\item At inner radii where the average surface density is higher than the stripping surface density, inhomogeneity leads to more gas at lower surface densities, allowing more stripping and faster acceleration of low density gas. We see this in the enhanced stripping rates in RC in the central regions (Fig. \ref{fig:surf_dens_W0_10}, \ref{fig:W0_10Myr} and \ref{fig:W0_35Myr}).

\item In the edge-on stripped SFW90, low-density galaxy fountain gas is stripped faster by the ICM wind (Fig. \ref{fig:W90_mass_flux}). However, this has a very small effect on the overall stripping rate, since the bulk of the gas is stripped in the galaxy plane. Instead, we find that the homogenised gas in SFW90 is easier to push at the beginning of RPS, thus allowing for the galaxy to be stripped faster (Fig. \ref{fig:W90_250Myr}). Once this lower density gas has been removed or shifted towards the centre of the galaxy (550 Myr), both RCW90 and SFW90 have almost identical stripping rates.

\end{enumerate}

We find that regardless of the wind impact angle stellar feedback has no direct effect on the stripping process, such as giving an additional velocity kick to the stripped gas. Instead, the indirect effect of homogenising the gas is what makes the difference between star-forming and radiative-cooling only galaxies.

An important insight we have gained from this comparison is that correctly modelling the small-scale distribution of gas in a disc undergoing RPS is important for understanding how quickly it will be stripped. We argue that cosmological simulations, which may find that feedback can enhance gas removal by RPS due to the strong ejection of gas particles to large distances from the galaxy plane \citep[e.g.][]{Bahe&McCarthy15, Kulier23}, cannot be used for detailed comparisons with RPS galaxy populations because they do not resolve the density range or distribution in the discs of satellite galaxies. We find that a more smooth gas distribution will result in more (compared to clumpy and underdense gas in RC simulation) stripping in regions with low average restoring force versus RP, and less stripping in regions with high average restoring force versus RP.

\section*{Acknowledgements}

This project has received funding from the European Research Council (ERC) under the European Union’s Horizon 2020 research and innovation programme (grant agreement No. 833824). The simulations were performed on the Frontera supercomputer operated by the Texas Advanced Computing Center (TACC) with LRAC allocation AST20007 and on the Stampede2 supercomputer operated by the Texas Advanced Computing Center (TACC) through allocation number TG-MCA06N030. B. V. acknowledges the Italian PRIN-Miur 2017 (PI A. Cimatti). We use the yt python package \citep{yt} for data analysis and visualisation.

%%%%%%%%%%%%%%%%%%%%%%%%%%%%%%%%%%%%%%%%%%%%%%%%%%
\section*{Data Availability}

The data underlying this article will be shared on reasonable request to the corresponding author.



\bibliographystyle{mnras}
\bibliography{GASP} % if your bibtex file is called example.bib

%%%%%%%%%%%%%%%%%%%%%%%%%%%%%%%%%%%%%%%%%%%%%%%%%%

%%%%%%%%%%%%%%%%% APPENDICES %%%%%%%%%%%%%%%%%%%%%

\appendix

\section{Refinement criteria and resolution test} \label{appendix:refinement}

\begin{figure*}
\centering
\includegraphics[width = 160mm]{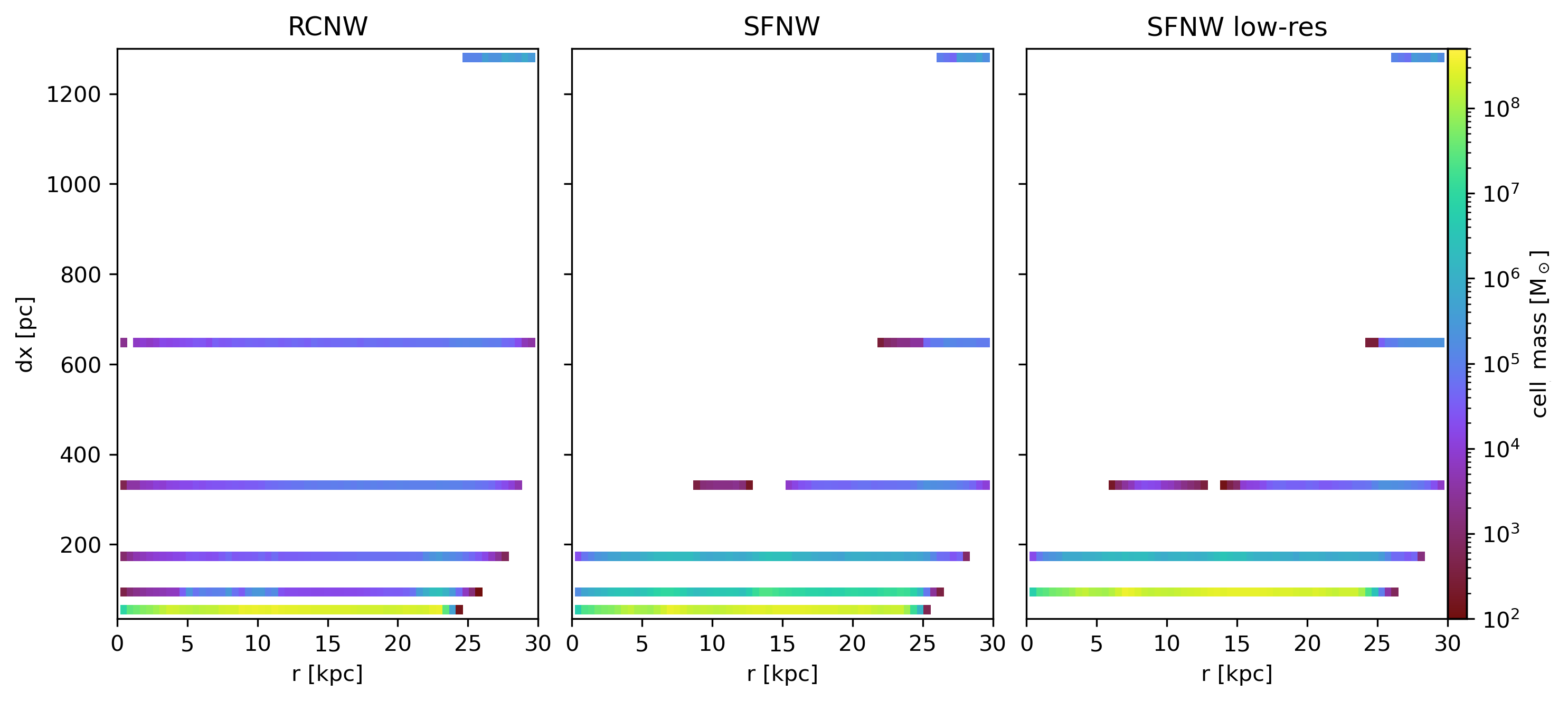}
\caption{Cell sizes $dx$ (we have cubical cells, $dx=dy=dz$) as a function of cylindrical radius $r$ colour-coded by cell mass in a galaxy disc ($R=30$ kpc and $z\pm2$ kpc) in RCNW (left) and SFNW (centre) and low-resolution SFNW (right) at 10 Myr \textit{before} RPS begins.}
\label{fig:refinement}
\end{figure*}

Our simulation box has 160 kpc on a side, divided into $128\times128\times128$ cells. We use the adaptive approach to include additional 5 levels of refinement, with each level having twice the resolution of the previous one, and get a maximum resolution of 39 pc. Our refinement criteria are Jeans length and baryon mass of $\approx7500 M_\odot$. In Fig. \ref{fig:refinement} we show how using these criteria we are able to refine the galaxy discs ($R=30$ kpc and $z\pm2$ kpc) in RCNW (left) and SFNW (centre) and low-resolution SFNW (right, see the text below). The cell sizes are colour-coded by cell mass. We select the data at 10 Myr \textit{before} RPS begins. Firstly, this plot shows that the chosen refinement criteria are able to highly resolve the disc, since in all of the galaxies the bulk of the mass is concentrated at the finest level. Secondly, compared to RCNW, SFNW has more mass on levels 4 and 5 (156 and 78 pc, respectively) since in this case the ISM disc within the $z\pm2$ kpc boundaries is much thicker (Fig. \ref{fig:NW_vz}).

Our galaxy, which is the best-resolved region in the simulation box, has a maximum of $3.33 \times 10^7$ of cells (of any resolution) in the SFNW run, while the median number of cells over the 1 Gyr of total simulation time is $2.86\times 10^7$. In the SF wind runs, during the first 100 Myr after the ICM has just hit the galaxies, the peak cell number value for different wind angles is $[3.36, 3.39] \times 10^7$. After that, the number of cells only continues to decrease with time, as the gas is being stripped and less and less regions are left to be resolved. In RCNW, as the lack of stellar feedback allows for the gas to collapse into a thin disc, the number of cells in the galaxy disc is somewhat smaller with the maximum and median values being $2.17 \times 10^7$ and $1.76 \times 10^7$, respectively.

Here we discuss how resolution may impact our results. We run a short simulation that includes only 4 levels of refinement, with the finest cell being 78 pc. In the central and right panels of Fig. \ref{fig:refinement} we can compare the resolution distribution in the fiducial and low-resolution runs, respectively. The bulk of the mass is located at the highest-resolution level, while the mass distribution on levels 1 to 4 (1248 to 156 pc) is similar in both simulations.

\begin{figure}
\centering
\includegraphics[width = 85mm]{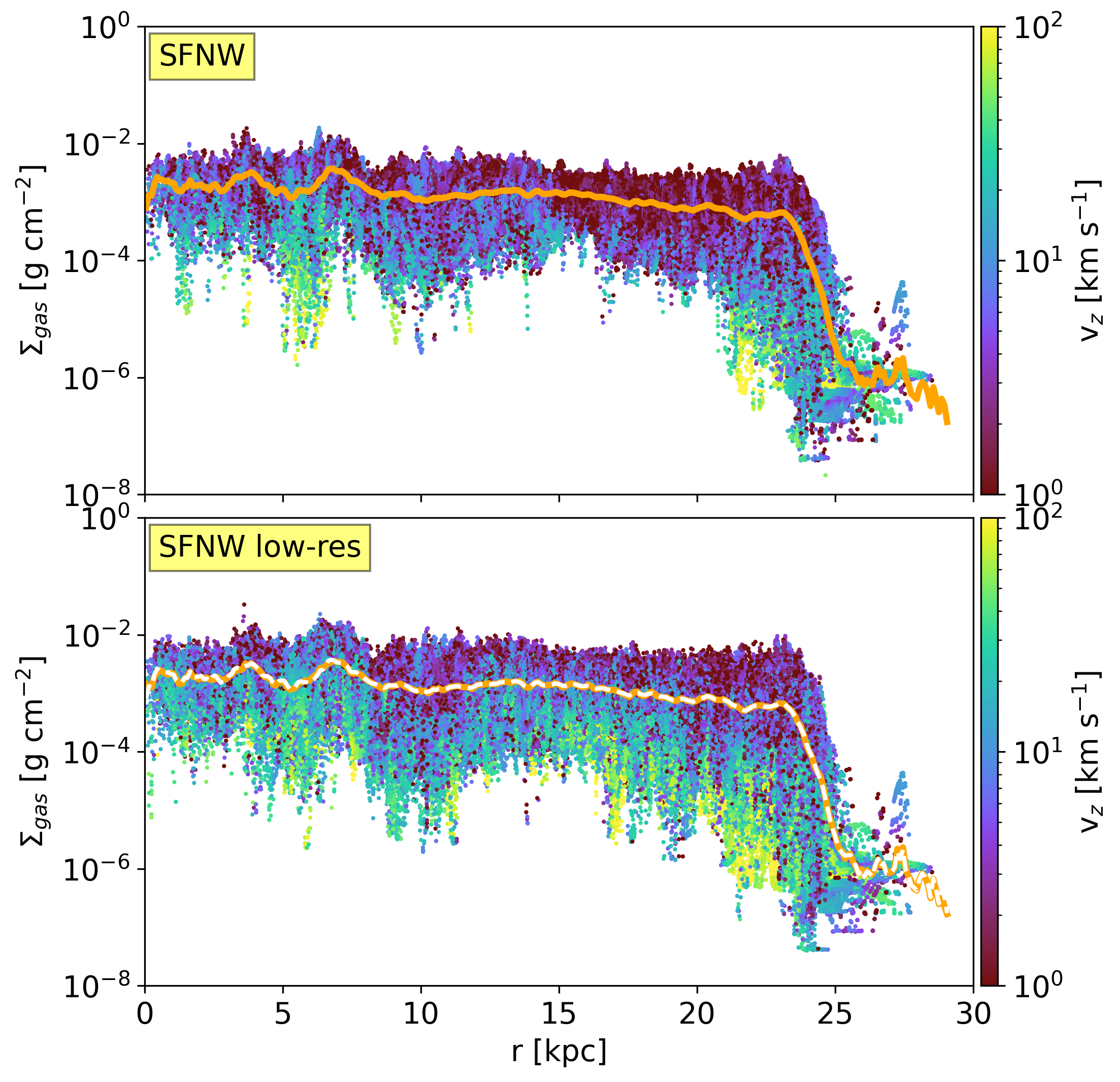}
\caption{Same as Fig. \ref{fig:surf_dens_W0_10}, but for the fiducial SFNW (top) and low-resolution (bottom) runs 10 Myr \textit{before} RPS begins. Note that the colour bar range is different in this figure, since before RP there is no high-velocity gas in the disc. In the bottom panel, the white dashed line overplots average surface density in the fiducial SFNW over the average in the low-resolution run to illustrate that they are almost identical.}
\label{fig:surf_dens_low_res}
\end{figure}

The main result of this paper, pictured in Fig. \ref{fig:surf_dens_W0_10}, is the role of the surface density distribution in gas stripping. The resolution used, therefore, must capture both the low- and high-density gas. In Fig. \ref{fig:surf_dens_low_res} we repeat Fig. \ref{fig:surf_dens_W0_10} for the fiducial SFNW (top) and low-resolution (bottom) runs 10 Myr \textit{before} RPS begins. Note that the colour bar range is different in this figure from Fig. \ref{fig:surf_dens_W0_10}, since before the RP, there is no high-velocity gas in the disc (in other words, we are missing gas that is being rapidly stripped). Overall, the surface density distribution in both galaxies looks quite similar, with some extension to lower surface densities in the SFNW low-res run. Meanwhile, the average surface densities are almost identical, as illustrated in the bottom panel with the white dashed line that overplots the average in the fiducial SFNW over the average in the low-resolution run. The similar average values mean that at any radius the bulk of the mass is distributed at comparable surface densities. We also note that this means that the calculated truncation radius (for which the average surface density is used) is resolution-independent.

\section{The effect of star formation on density distribution} \label{appendix:star-formation}

In this work we compared radiative-cooling-only simulations to the ones that include star formation and stellar feedback. Here, we will consider the effect that star formation alone has on the gas distribution in the galaxy. This will help us separate the role of star formation from that of stellar feedback in shaping the ISM.

We will start by outlining the findings of previous works that used, like us, the Enzo code and considered the impact of star formation and feedback on the ISM. For example, \cite{TaskerBryan06} run a suite of hydrodynamical simulations of a Milky-Way-like galaxy. They study the effect that star formation threshold density and star formation efficiency play in shaping the Kennicutt-Schmidt law, and run radiative-cooling-only, star-formation-only and simulations that include both star formation and stellar-feedback. Comparison of the density-temperature gas distribution between their star-formation-only and feedback-including simulations (their Figs 11 and 12) reveals that the low-density-high-temperature ISM is produced (for the most part) and is most strongly affected by the stellar feedback. This is evident from both the mass and the volume distributions. Conversely, the high-density-low-temperature ISM (`molecular' or star-forming gas) is affected mostly by star formation, although some density redistribution by stellar feedback is also evident. \cite{Goldbaum16} also makes a comparison between star-formation-only and feedback-including simulations. Although this is not the focus of their work, they also show that it is the feedback that is responsible for smoothing the density distribution of the multi-phase ISM by preventing much of the mass from being bound into dense clouds. Therefore, star formation alone should not cause the lack of the lowest surface density gas found when comparing SFW0 to RCW0 in Fig. \ref{fig:surf_dens_W0_10}.

\begin{figure}
\centering
\includegraphics[width = 85mm]{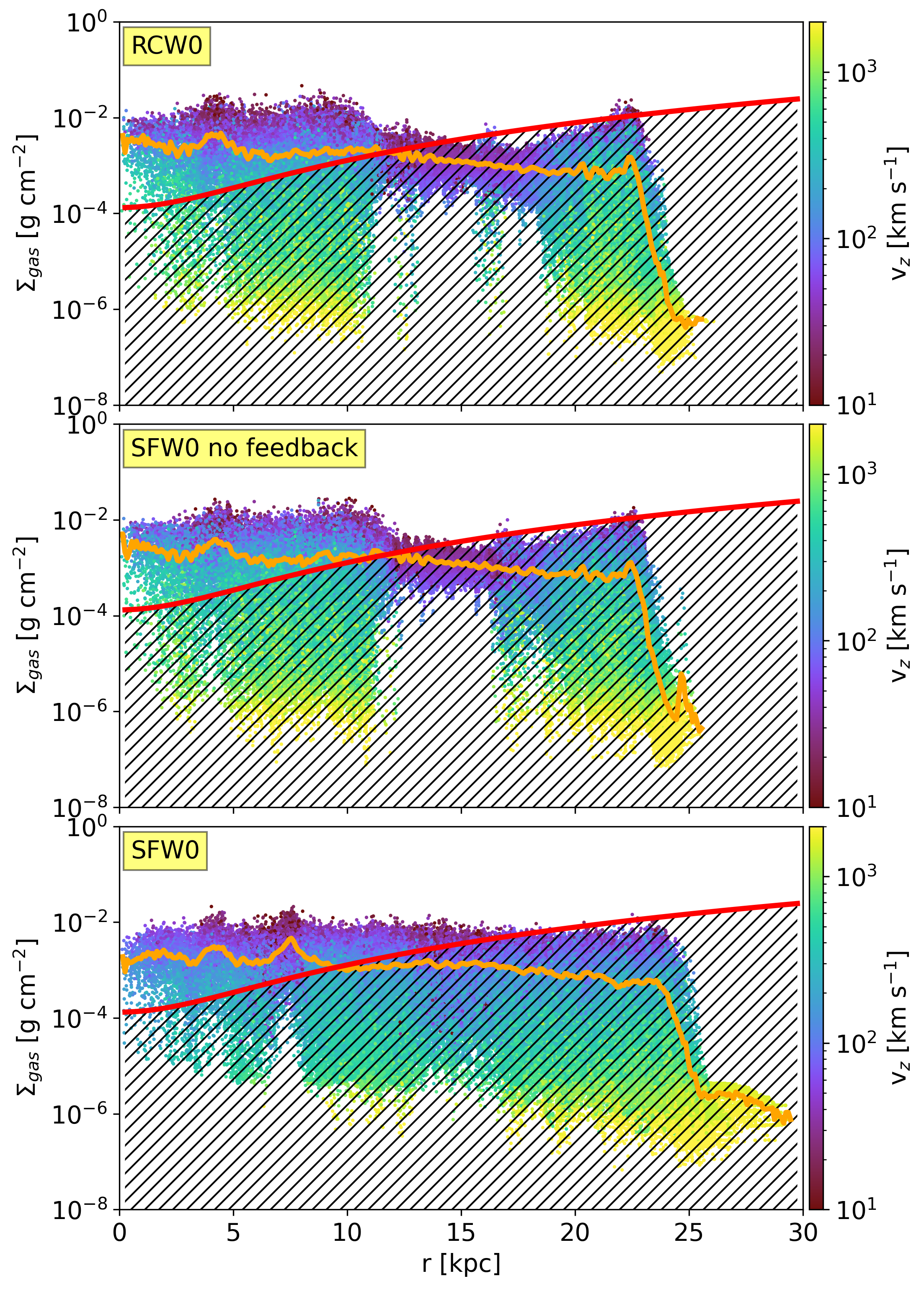}
\caption{Same as Fig. \ref{fig:surf_dens_W0_10}, but for RCWO (top) SFW0-no-stellar-feedback (middle) and SFW0 (bottom) runs at 10 Myr.}
\label{fig:surf_dens_noFeedback}
\end{figure}

To confirm these previous works, we run a short simulation similar to SFW0 but without stellar feedback. In order to save computational resources, we restart RCW0 40 Myr prior to RPS (160 Myr from the beginning of the simulation) and turn on only the star formation with the same parameters as in our fiducial model. In Fig. \ref{fig:surf_dens_noFeedback} we repeat the Fig. \ref{fig:surf_dens_W0_10} but for RCW0 (top) SFW0-no-stellar-feedback (middle) and SFW0 (bottom) galaxies (50 Myr later, 10 Myr after the ICM wind has hit the galaxy). As expected from previous studies, the star formation alone is not enough to affect the ISM, especially its low-density phase, and, in fact, the only noticeable difference from RCW0 is in slightly lower density peaks as the dense gas can now form stars. The run with star formation and feedback (bottom panel), however, has the lowest density peaks and, particularly in the central regions, has a narrower range in gas surface density. We conclude that it is the stellar feedback that creates the more homogeneous multi-phase ISM in SFW0.

Another factor that could potentially influence our results is the value of the star formation threshold density. As we explained in Section \ref{sec:methods}, according to the chosen star formation recipe, if the mass of a cell exceeds the Jeans mass and the minimum threshold number density, a stellar particle will form with a minimum mass of $10^3 M_\odot$, with a chosen star formation efficiency of 1 per cent. In the fiducial SFNW (in this Appendix referred to as SF10) we choose a threshold number density of $n_{\rm{thresh}}=10 \, \rm{cm}^{-3}$. We also simulate galaxies from the same initial conditions with $n_{\rm{thresh}} = 3 \, \rm{cm}^{-3}$ (SF03) and $30 \, \rm{cm}^{-3}$ (SF30) for 300 Myr (the time at which the ICM wind hits the galaxies in the fiducial star-forming runs). 

\begin{figure}
\centering
\includegraphics[width = 85mm]{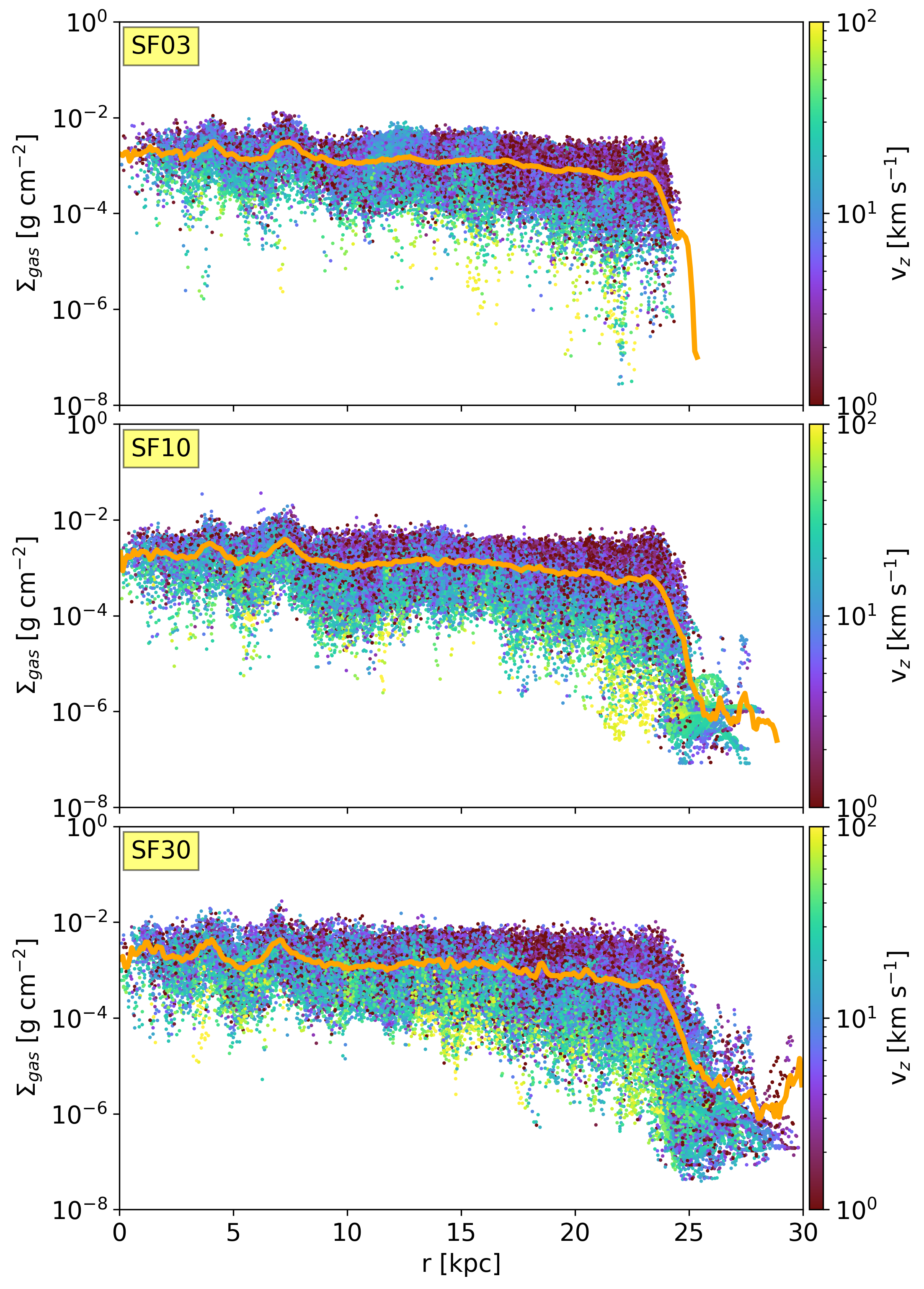}
\caption{Same as Fig. \ref{fig:surf_dens_W0_10}, but for the SF03 ($n_{\rm{thresh}} = 3 \, \rm{cm}^{-3}$, top), SF10 (or SFNW, $n_{\rm{thresh}} = 10 \, \rm{cm}^{-3}$, middle) and SF30 ($n_{\rm{thresh}} = 30 \, \rm{cm}^{-3}$, bottom) runs at 300 Myr of {\it total} evolution, at which point the wind would hit the galaxies in the fiducial runs. Note that the colour bar range is different in this figure, since without RP there is no high-velocity gas in the disc.}
\label{fig:surf_dens_SF}
\end{figure}

\begin{figure}
\centering
\includegraphics[width = 70mm]{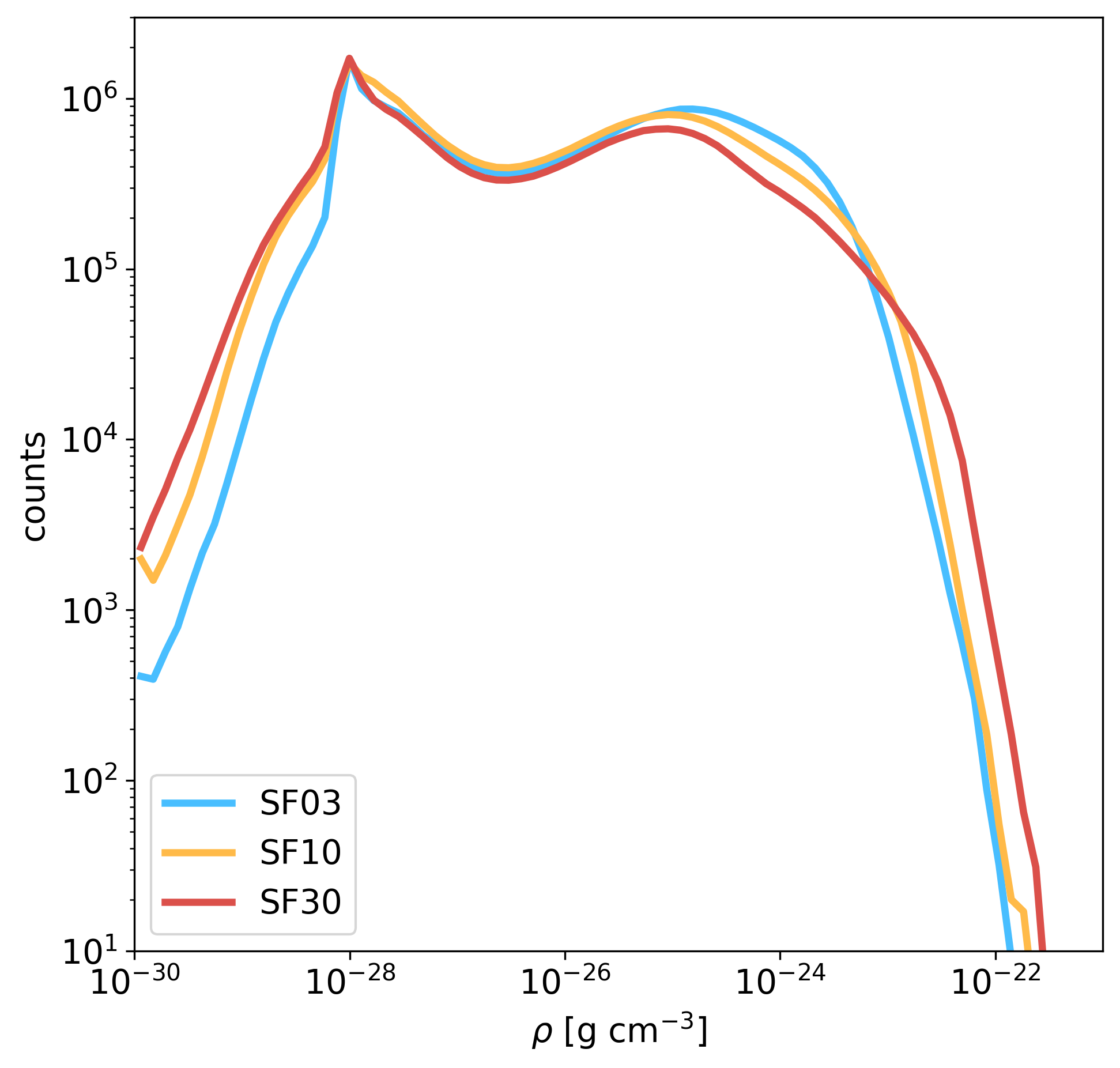}
\caption{For SF03 ($n_{\rm{thresh}} = 3 \, \rm{cm}^{-3}$, blue), SF10 (or SFNW, $n_{\rm{thresh}} = 10 \, \rm{cm}^{-3}$, yellow) and SF30 ($n_{\rm{thresh}} = 30 \, \rm{cm}^{-3}$, red) runs, the volume density histogram of gas ($Z > 0.35 Z_\odot$) in a galaxy disc ($R=30$ kpc and $z\pm2$ kpc) taken at 300 Myr of {\it total} evolution.}
\label{fig:dens_hist}
\end{figure}

We find that the differences in the ISM brought by the change of threshold number density are overall minor. We illustrate this by once again repeating Fig. \ref{fig:surf_dens_W0_10}. In Fig. \ref{fig:surf_dens_SF} we plot surface density distribution, this time for the SF03 (top), SF10 (middle) and SF30 (bottom) runs. The central regions, where the bulk of star formation occurs, are quite similar in SFR and gas surface density across all the density thresholds. The gas surface density average and distributions are very similar in all cases until at the very edge of the SF03 disc where there is little ISM gas as it has been used in star formation. The differences in the ISM brought by the change of threshold number density are overall unsubstantial. We further prove this by plotting in Fig. \ref{fig:dens_hist} the volume density histograms of all the cells in the ISM of the galaxy disc ($R=30$ kpc and $z\pm2$ kpc) for SF03 (blue), SF10 (yellow) and SF30 (red) runs, which are very close to each other. Thus, we conclude that the choice of the number density threshold for star formation would not have a significant influence over our results.

\section{Gas fallback in face-on stripped galaxies} \label{appendix:negative}

\begin{figure*}
\centering
\includegraphics[width = 175mm]{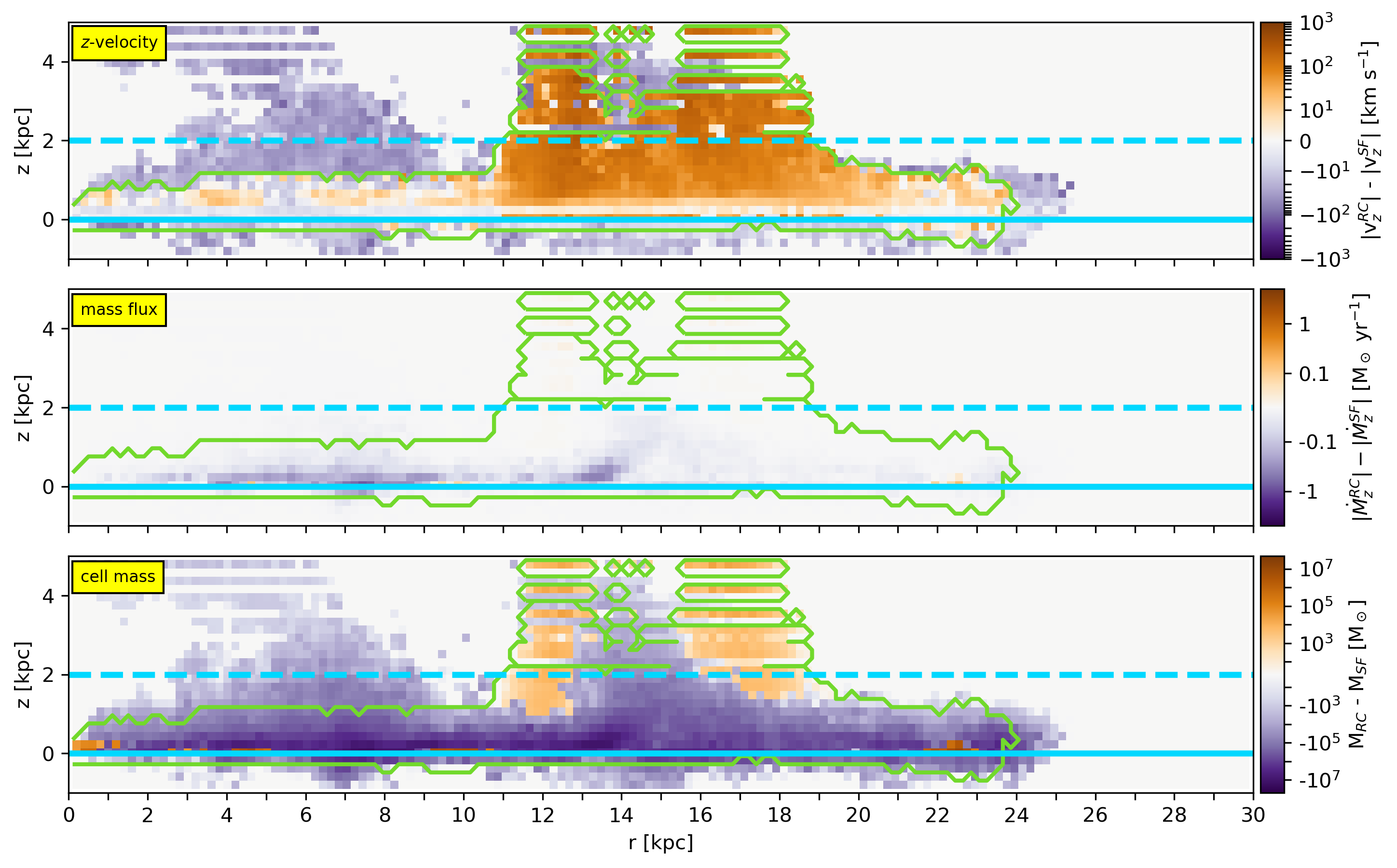}
\caption{Same as Fig. \ref{fig:W0_10Myr} but for $v_z<0.$}
\label{fig:W0_10Myr_negative}
\end{figure*}

\begin{figure*}
\centering
\includegraphics[width = 175mm]{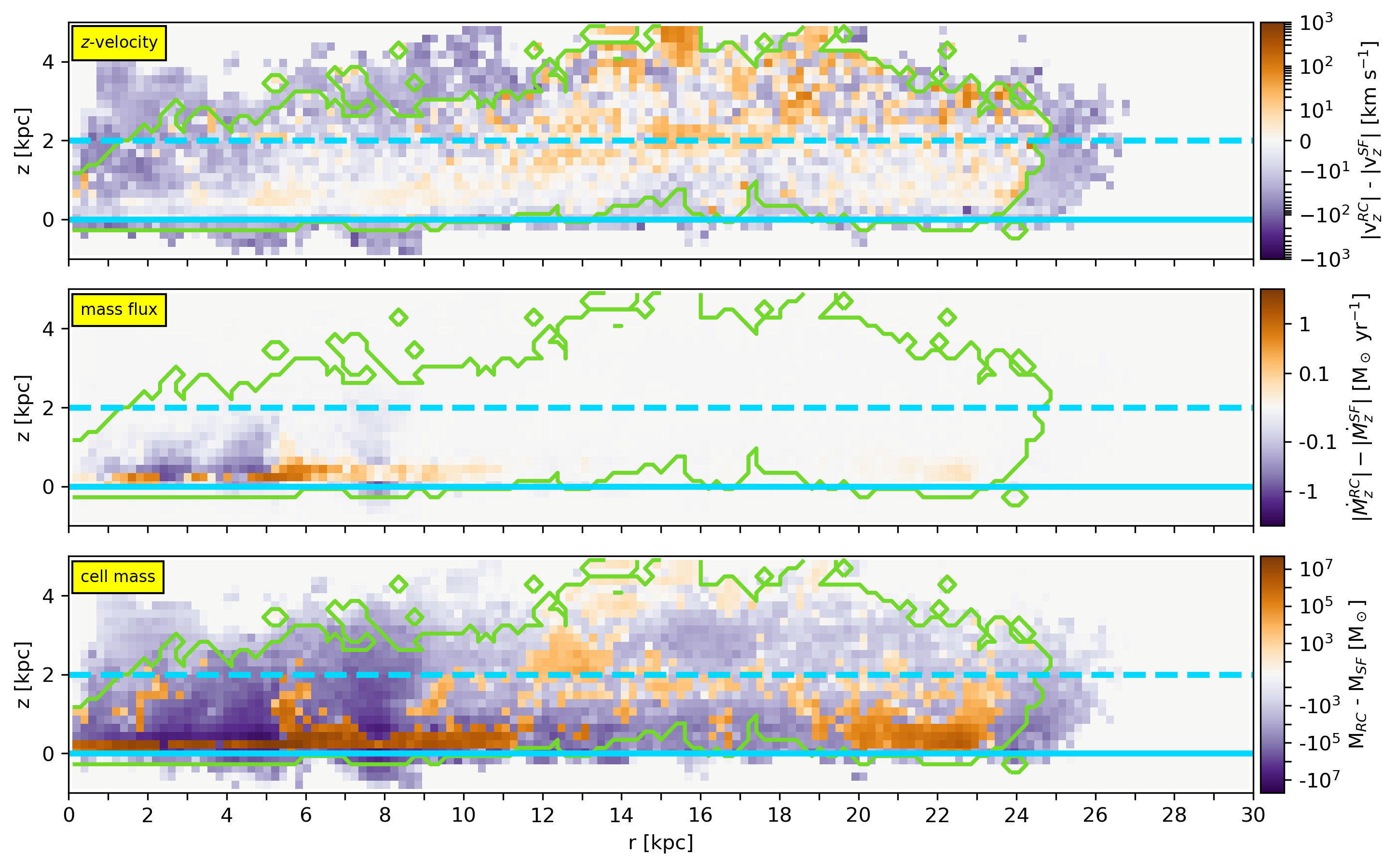}
\caption{Same as Fig. \ref{fig:W0_35Myr} but for $v_z<0.$}
\label{fig:W0_35Myr_negative}
\end{figure*}

Here for completeness we show the gas fallback in the face-on stripped galaxies. We repeat the plots from Section \ref{subsec:face-on} but for $v_z < 0$ so that Fig. \ref{fig:W0_10Myr_negative} is for 10 Myr (same as Fig. \ref{fig:W0_10Myr}) and Fig. \ref{fig:W0_35Myr_negative} is for 35 Myr (same as Fig. \ref{fig:W0_35Myr}). To avoid confusion, we show the difference of {\it absolute} values of velocity (top panels) and mass flux (middle panels), so that, like in the previous plots, orange colour means faster gas, more mass flux in RCW0 and purple colour means that SFW0 dominates.

In \cite{Akerman23} we show that fallback is negligible compared to the radial flows withing the galaxy when it is subject to a varying, ever-increasing ICM wind. It is no surprise, then, to see that on the maps at both 10 and 35 Myr the mass flux is almost 0 (notice that the colour bar scale is the same as in Fig. with $v_z > 0$). Since in green we outline the gas distribution in RCW0, we can easily compare it to Fig. \ref{fig:W0_10Myr} and \ref{fig:W0_35Myr} and find that the gas with negative $v_z$ is several kpc less radially extended than the gas with positive $v_z$. This means that there is no fallback in the outskirts or in the regions with the most violent stripping. 

We can only see some fallback in the inner 10 kpc where it is dominated by SFW0. As explained in Section \ref{subsec:face-on}, in SFW0 this part of the galaxy is protected from RPS, since it is located within the stripping radius and the outflows driven by feedback are able to fall back to the galaxy in fountain flows. On the other hand, through the big low-density holes in the RCW0 gas distribution the ICM wind can move freely, dragging some of the ICM with it. As we can see here, this gas is truly stripped and does not fall back on the galaxy.

%%%%%%%%%%%%%%%%%%%%%%%%%%%%%%%%%%%%%%%%%%%%%%%%%%

% Don't change these lines
\bsp	% typesetting comment
\label{lastpage}
\end{document}